\documentclass[letterpaper,oneside,10pt,twocolumn]{article}

\usepackage[T1]{fontenc}

\usepackage{abstract}

\usepackage{authblk}

\usepackage{floatrow}
\usepackage[toc,page]{appendix}
\usepackage{xcolor}
\usepackage{lmodern}
\usepackage{ifthen}
\usepackage{cite}
\usepackage[colorlinks,citecolor=cyan,backref=page]{hyperref}
\renewcommand*{\backrefalt}[4]{%
  \ifcase #1 %
    No citations.%
  \or
    See p.~#2.%
  \else
    See pp.~#2.%
  \fi
}

\usepackage{amssymb}
\usepackage{fancyhdr}
\usepackage{lastpage}
\usepackage{amsmath}
\usepackage{empheq}

\usepackage{lineno}

\usepackage{geometry} 
\fancypagestyle{plain}{%
  \fancyhf{}%
  \fancyhead[R]{FERMILAB-FN-1061-ND}
  \fancyfoot[C]{\thepage\ of \pageref*{LastPage}}}

\usepackage{array}
\usepackage{multirow}
\usepackage{rotating, graphicx}
\usepackage{xspace}
\usepackage{verbatim}
\usepackage{csquotes}

\newcommand{\nova}{NOvA\xspace}

\newcommand{\ul}{\begin{itemize}}
\newcommand{\lu}{\end{itemize}}
\newcommand{\ol}{\begin{enumerate}}
\newcommand{\lo}{\end{enumerate}}

\newcommand{\is}[2]{\texorpdfstring{${}^{#1}$#2}{#2-#1}}

\newcommand{\dedx}{$\mathrm{d}E/\mathrm{d}x$\xspace}
\newcommand{\piz}{$\pi^0$\xspace}
\newcommand{\pip}{$\pi^+$\xspace}
\newcommand{\pim}{$\pi^-$\xspace}
\newcommand{\mup}{$\mu^+$\xspace}

\newcommand{\mum}{\texorpdfstring{$\mu^-$\xspace}{mu\xspace}}

\newcommand{\fig}[1]{Fig.~\ref{fig:#1}}

\newcommand{\tab}[1]{Table~\ref{tab:#1}}

\newcommand{\rf}[1]{Ref.~\cite{#1}}
\newcommand{\rfs}[1]{Refs.~\cite{#1}}

\newcommand{\sect}[1]{Section~\ref{sec:#1}}
\newcommand{\sects}[1]{Sections~\ref{#1}}

\newcommand{\steeltext}{
The error on the mass itself ranges from \steelrawmasserrormin to
\steelrawmasserrormax, depending on the number of planes considered. Since
muons are only selected if they enter from the main detector, only consecutive
combinations of plane starting with the upstream end are relevant.  ``Mass''
here means the product of density and thickness in the region away from the
edges of the steel planes. The error is predominantly due to the thickness;
the density is measured with significantly higher precision.

Non-uniformity in the steel introduces additional uncertainty since muon tracks
do not sample the steel uniformly.  The steel has thickness variations on order
of 1--2\,mm, but there does not seem to be any systematic trend (e.g. the top
being thicker than the bottom), so the effects of this should mostly cancel.
If the variations were purely random and uncorrelated from point to point, no
additional uncertainty for the muon range would be needed even given
non-uniform sampling.  To account for the possibility that there is some
significant correlation, a simple Monte Carlo simulation has been performed.
Steel planes are constructed with a random distribution of thicknesses across
their surfaces drawn from the observed distributions.  Many such sets of steel
plates are constructed, with the length scale of the thickness variations
chosen randomly, on a log scale, between 1\,cm and 1\,m.  Muons with a
realistic distribution of initial positions and angles are propagated through
them and the thicknesses of steel encountered by each is recorded provided
that the muon stays contained within the Muon Catcher.  The
distribution of deviations between the mean thickness of steel encountered by
these muons and the nominal mean steel thickness gives the non-uniformity
error.  The effect is found to be very small, with the RMS of these deviations
ranging from \steeluniformityerrormin to \steeluniformityerrormax, depending on
the number of planes that a muon intersects.

The steel planes are painted to prevent rust.  The very small mass of the paint
has not been measured, but using the typical thicknesses of coats of paint, a
\steelpaintmasserror uncertainty is assigned to the steel plane mass to account
for it. 

The summed error for the steel planes (mass, uniformity and paint) ranges from
\steelmasserrormin to \steelmasserror, depending on the number of planes that a
muon intersects, with a mean of \steelmasserrormean.  Since the mean and
maximum errors are similar, the maximum is adopted as the error for the whole
Muon Catcher from these effects.
}

\newcommand{\coulombsection}[1]{

\section{Coulomb corrections}\label{sec:coulomb}

The energy of the outgoing muon in a neutrino-nucleus interaction is modified
by the Coulomb field of the nucleus.  For \mum, the muon is attracted to
the nucleus, so energy is lost, and vice versa for \mup.
There exists an implementation of this effect in \genie, but it is disabled
by default, and is left disabled for NOvA.  This is just as well, because
the version currently used by NOvA, 2.12.10b, has a bug in which it subtracts
energy from the lepton regardless of charge.  \ifthenelse{#1 = 0}{I have
submitted a pull request to fix
this.\footnote{\texttt{\href{https://github.com/GENIE-MC/Generator/pull/5}{https://github.com/GENIE-MC/Generator/pull/5}}}}\xspace

A simple form of the correction, as implemented by \genie, is~\cite{minosdoc3172,Aste2005}:
\[R = (1.1A^{1/3} + 0.86A^{-1/3})/\hbar c\]
\[\Delta E = 0.75 \times 3Z/2\alpha R, \]
where $\alpha$ is the fine structure constant.
The mean effect in NOvA for $\nu_\mu$ interactions,
which create \mum, is $-$\coulombcorrmuminusmevnonothing\,MeV, and for $\bar\nu_\mu$ interactions,
which create \mup, is $+$\coulombcorrmuplusmevnonothing\,MeV.  See \tab{coulomb}. The asymmetry is caused
by the charge of the residual nucleus, which differs by 2 units, including the
fact that $\bar\nu_\mu$
can interact on hydrogen to produce a neutron, for which there is no correction.

\begin{table}

\caption{\label{tab:coulomb} Coulomb corrections for \mum and \mup, and the rough fraction
of (anti)neutrino interactions on each element.}

\begin{center}
\begin{tabular}{c c c | c c}

\hline
\hline
& \mum shift (MeV) & Fraction & \mup shift (MeV) & Fraction \\
\hline

H & $-$\coulombcorrmuminusmevH & \coulombcorrmuminuspercentH & 0 & \coulombcorrmupluspercentH \\
C & $-$\coulombcorrmuminusmevC & \coulombcorrmuminuspercentC & $+$\coulombcorrmuplusmevC & \coulombcorrmupluspercentC \\
O & $-$\coulombcorrmuminusmevO & \coulombcorrmuminuspercentO & $+$\coulombcorrmuplusmevO & \coulombcorrmupluspercentO \\
Cl & $-$\coulombcorrmuminusmevCl & \coulombcorrmuminuspercentCl & $+$\coulombcorrmuplusmevCl & \coulombcorrmupluspercentCl \\
Ti & $-$\coulombcorrmuminusmevTi & \coulombcorrmuminuspercentTi & $+$\coulombcorrmuplusmevTi & \coulombcorrmupluspercentTi \\
Sn & $-$\coulombcorrmuminusmevSn & \coulombcorrmuminuspercentSn & $+$\coulombcorrmuplusmevSn & \coulombcorrmupluspercentSn \\

\hline
\hline

\end{tabular}
\end{center}
\end{table}

Bodek~\cite{Bodek:2018lmc} gives a form with somewhat different coefficients, yielding results
7\% smaller.  It seems reasonable, then, to take the correction to be accurate to 10\%.
At 1\,GeV,
this is \coulombcorrmuminuserror for \mum and \coulombcorrmupluserror for \mup.
It's notable that Bodek also recommends a second correction for \genie to the outgoing
lepton energy.  Neither this correction nor any uncertainty on it has been used thus
far by NOvA.  Ultimately, it may be the correct approach to fold both corrections into
a single unified treatment that ensures MeV-level energy conservation.

\ifthenelse{#1 = 0}{
In order to use these small errors, we must repair the central values in the
energy estimation code.  The best solution for this would be to fix \genie, enable
the correction, make a new production, and retune the energy estimator.  Eventually,
we can do all that.  We discussed putting in a switch to the muon energy estimation
code which, if the event is data and FHC (RHC) adds \coulombcorrmuminusmevnonothing\,MeV
(subtracts \coulombcorrmuplusmevnonothing\,MeV) and if it is MC
does nothing.  But then we decided to do nothing for the 2019 analysis that would shift
the central value gotten from data already analyzed in 2018. 

At our 2018-10-04 ``NOvA weekly alternating $\nu_\mu$/$\nu_\mathrm e$ meeting
with a $\nu_\mu$ focus'', we decided that the Coulomb and Bodek corrections
would be handled as systematic errors separate from the muon range error
discussed in this note. 

}{
It has been decided to treat these concerns separately from the muon range error in 
future NOvA analyses.
}

}

\newcommand{\dedxtable}{
\begin{table}
\begin{center}
\begin{tabular}{l c}

\hline
\hline

Material    & \dedx (MeV/cm$^2$/g) \\

\hline
Scintillator & 2.083 \\
Glue & 1.943 \\
Air & 1.809 \\
PVC & 1.767 \\
Steel & 1.452 \\

\hline
\hline
\end{tabular}
\end{center}
\caption{\label{tab:dedx} Nominal \dedx for each material at minimum ionization, in decreasing
order.}
\end{table}
}

\newcommand{\massaccountingtable}{
\begin{table}
\begin{center}
\noindent \begin{tabular}{l c c c}
\hline
\hline
Component & Mass & \dedx & Principle cause \\
\hline
FD scintillator & \fdscinterror   & & Volume \\
ND scintillator & \ndscinterror   & & Density, volume \\
FD PVC          & \fdpvcerror     & & Weighing \\
ND PVC          & \ndpvcerror     & & Weighing \\
Glue (both)     & \glueerror      & & Outgassing \\
Steel           & \steelmasserror & & Volume \\
\hline
FD           & \fdmasserror          & \fddedxmasserror & PVC \\
Main ND      & \ndmasserror          & \nddedxmasserror & PVC \\
Muon catcher & \muoncatchermasserror & \muoncatcherdedxmasserror & Steel \\
\hline
Main ND/FD & \nearfarmasserror & \nearfardedxmasserror & FD scintillator \\
~ ~ ~ \emph{From scintillator} & & \emph{\ndfdscinterrorcontrib} &  \\
~ ~ ~ \emph{From PVC} & & \emph{\ndfdpvcerrorcontrib} &  \\
Muon catcher/FD & \mufdratiomasserror & \mufdratiodedxmasserror & Steel \\
~ ~ ~ \emph{From steel} & & \emph{\mufdsteelerrorcontrib} & \\
Muon catcher/main ND & \mundratiomasserror & \mundratiodedxmasserror & Steel \\
~ ~ ~ \emph{From steel} & & \emph{\mutoactivesteelerrorcontrib} & \\
\hline
\hline
\end{tabular}
\end{center}
\caption{\label{tab:mass} Mass accounting errors.  The second column shows the
raw mass errors, while, for whole detectors, the third column shows
the combinations weighted by \dedx.  Those shown in the
top section for the individual materials are relative to the total mass
of each of those materials alone. The remaining errors are combinations
of these weighted by each materials' nominal \dedx.}
\end{table}
}

\newcommand{\pdffigure}{
\begin{figure}
\begin{center}
\includegraphics[width=\columnwidth]{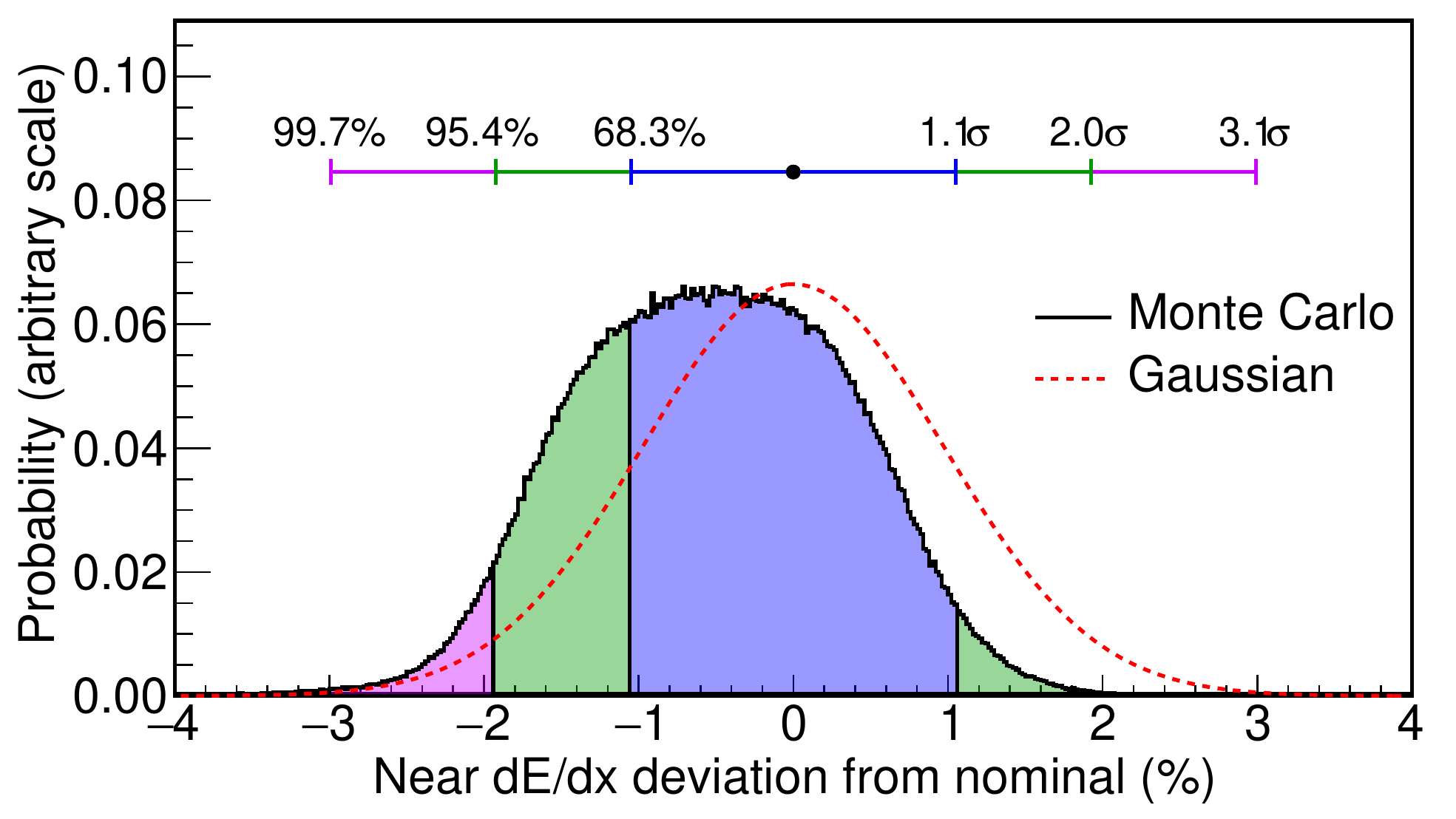}

\includegraphics[width=\columnwidth]{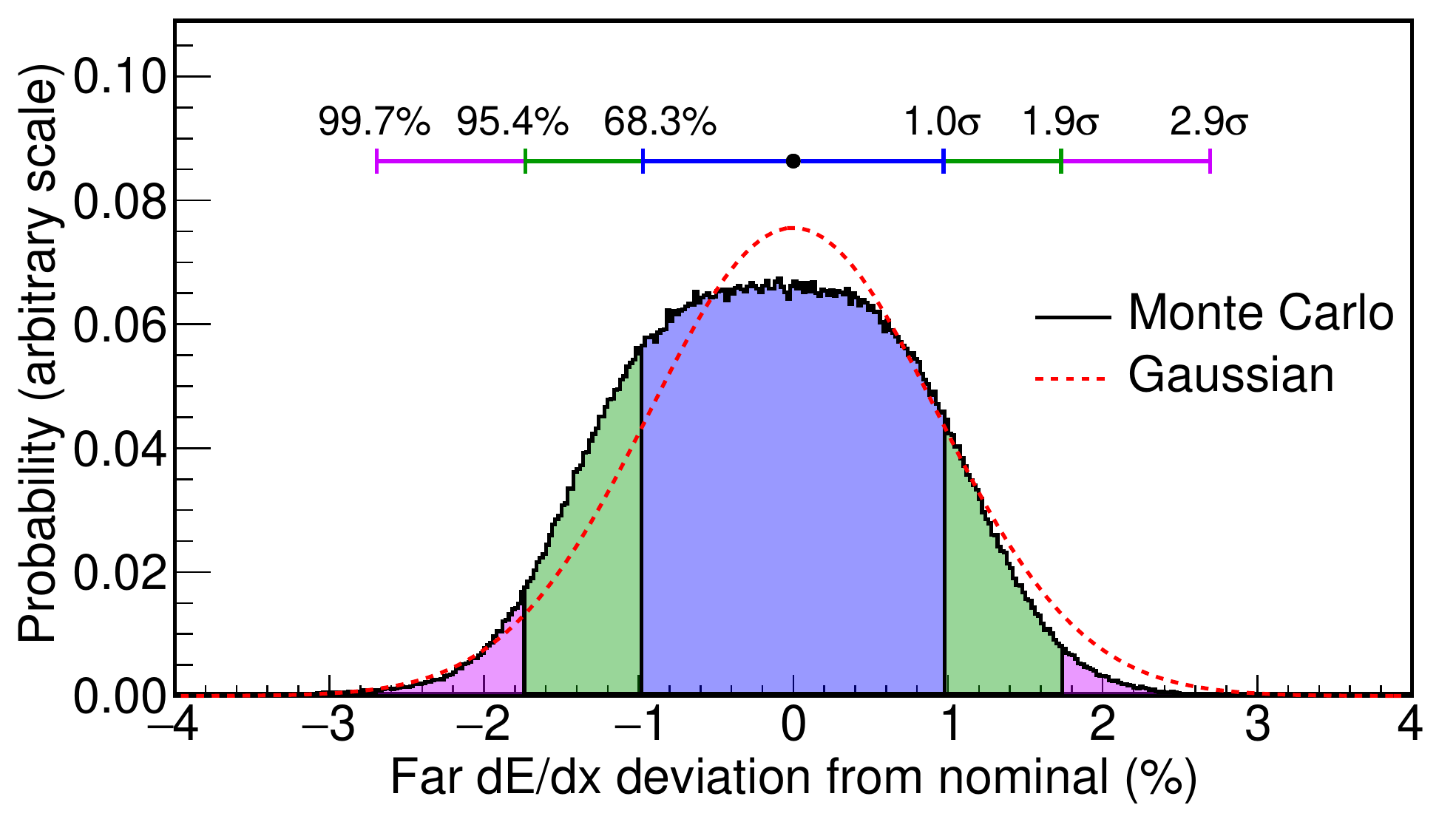}

\includegraphics[width=\columnwidth]{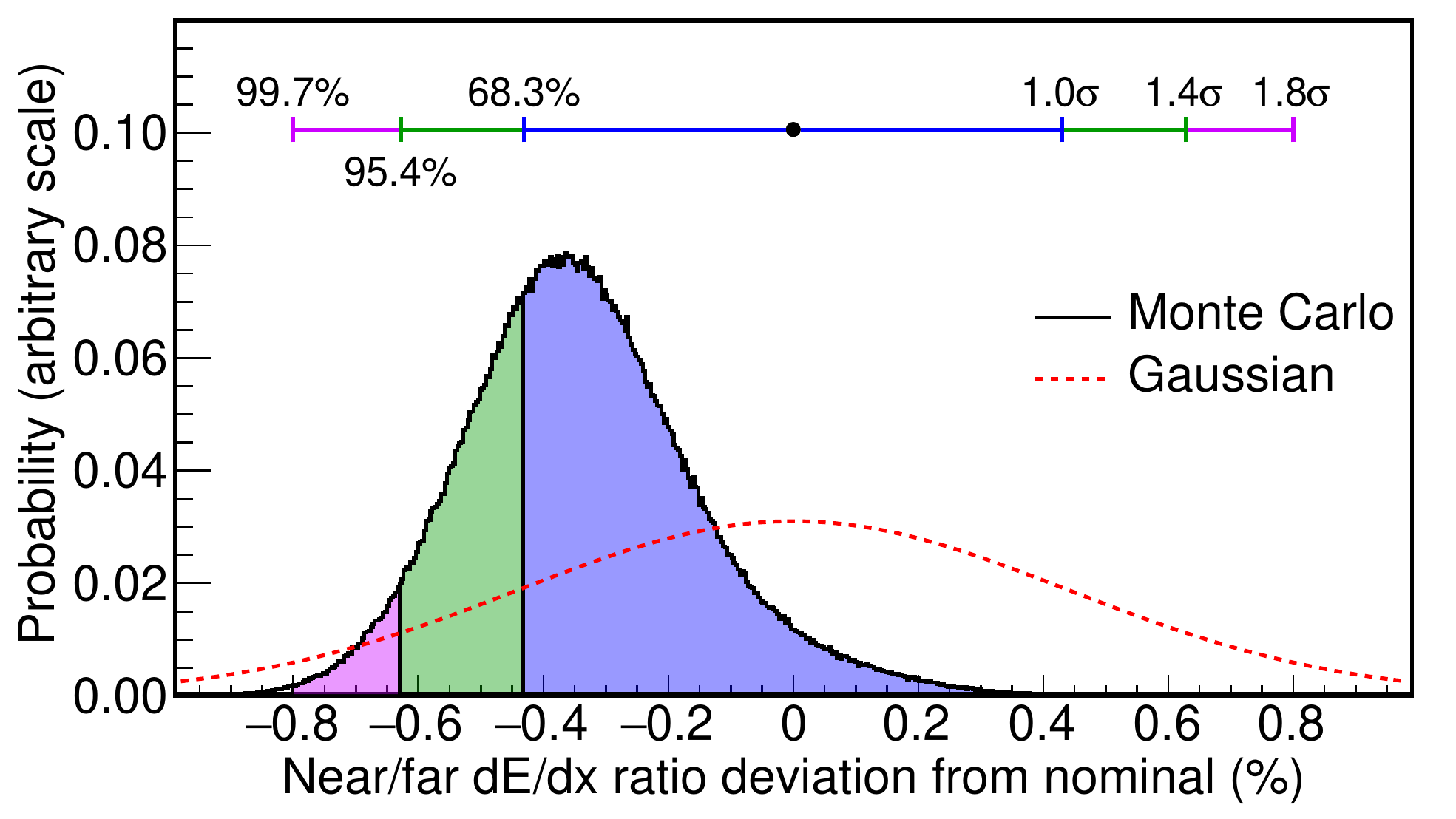}
\end{center}
\caption{\label{fig:nearfar} Comparison of a PDF with non-Gaussian components
for the Near (top), Far (middle), and Far/Near ratio (bottom) errors,  and the
Gaussians with approximating them that are used in the analysis.  The colors
and horizontal error bars show the extents of centered regions containing
68.27\%, 95.45\% and 99.73\% of the probability, and how many ``$\sigma$''
(units of the nominal error used in the Gaussian approximation) they contain.
When these are smaller than \{1,2,3\}$\sigma$, it indicates that use of the
Gaussian approximation is conservative. Asymmetries primarily indicate that our
best estimate includes a shift relative to the nominal detector Monte Carlo,
but because implementation of shifts are error-prone, a centered error has been
used to cover the shift.}
\end{figure}
}

\newcommand{\muoncatcherneutrontext}[1]{

Over the course of a 10\,$\mu$s beam
spill, many thermalized neutrino-induced neutrons build up in the ND,
some produced in the detector, and some in the surrounding rock.  As the
neutron capture time in the detector is about 50\,$\mu$s, the number of neutron
captures, most on \is{35}Cl producing 8.6\,MeV of gamma rays, increases roughly
linearly over the spill.  Typically these capture events produce one hit that
cannot be readily connected to its parent neutrino event, since it is well
separated in time, and neutrons can travel long distances before being
captured.  This effect is negligible at the FD, which has much less than one
neutrino interaction per beam spill on average.

These stray neutron hits can randomly appear near the beginning or end of muon tracks,
causing the reconstructed track length to be increased.  It has been found
that, even though neutron thermalization and capture is modeled by our MC, this
effect is much more prominent  in data than in MC.  The causes of this are
still under investigation, and may have to do with neutron modeling, details of
hit timing, or both.

The effect is most clear in the Muon Catcher. For full-intensity beam spills,
tracks ending in the Muon Catcher near the end of the spill are reconstructed
with $17\pm5$\,MeV more energy on average than those at the beginning. \ifthenelse{#1 = 0}{(It is not
yet definitive that this increase is caused by neutrons, but \emph{something}
is skewing the reconstruction as a function of time, which is an effect that
needs to be covered.)}\xspace On the
other hand, the effect in the main ND is too small to clearly measure, with an
upper limit of perhaps 10\,MeV\ifthenelse{#1 = 0}{~\cite{novadoc31144}}.

This difference is understood as the combination of two geometric effects.  First, pairs of
planes in the Muon Catcher are separated by 10\,cm of steel, so a randomly
placed hit added to a track typically adds a much greater amount of material
that the muon is incorrectly inferred to have traveled through.  Second, for a
fixed angular acceptance, the tracker has more cells that it will accept as
part of a track in the Muon Catcher since the active planes are offset by steel
planes.  These together give about a factor of 10.  This is reduced by perhaps
a factor of two by the fact that many of the neutron-produced gammas in the
muon catcher are absorbed by the steel.

A \muneutronpileuperrormev\,MeV systematic error is taken for the Muon Catcher
to cover the mean of the effect for full intensity spills.  This is a bit less than
half of the difference in observed track energies to account because it is
assumed that part of the effect occurs at the upstream end of the track in the
main ND.  For the main ND, an error of
\ndneutronpileuperrormev\,MeV is used. These errors are most applicable to full
intensity beam in neutrino mode.  In anti-neutrino mode, the number of
interactions per spill is lower by about a factor of three, reducing the effect
roughly in proportion.

Because the length of selected muon tracks in the Muon Catcher can be very
short, it is awkward to represent the error as a single percentage. As shown in
\fig{ucplot}, the effect of neutron pileup (along with other much smaller
energy-independent effects) can be as much as \mucatchertotalerror.  \ifthenelse{#1 = 0}{Assuming we want to keep
the implementation of the error very simple, I recommend a flat error of \mucatchertotalerrorabs\,MeV,
which is the maximum absolute error and overestimates the error at low energy
by a factor of three, rather than a flat error of \mucatchertotalerror, which is the maximum relative
error and overestimates the error at high energy by a factor of five.}
{
To keep the implementation of the error simple, instead a flat error of \mucatchertotalerrorabs\,MeV
is used.  This is the maximum absolute error and overestimates the error in the Muon Catcher
for the shortest tracks by a factor of three.  In future analyses, the pile-up error may be considered separately
to avoid this overestimate.
}

}

\newcommand{\summarytable}{

\begin{small}
\begin{center}
\begin{tabular}{l c c | c | c | c | c | c | c | c c}
\hline
\hline
       & \!\!\!$E_\mathrm{ref}$\,(GeV)\!\!\! & Mass & $I$ & Compounds & $\delta$ & Pile-up & Hadronic & \textbf{Total} & Dominant & Next \\
\hline
\multicolumn{3}{l|}{Absolute}         &                   &                           &                     &                       &                           &                               & \\
FD&\referenceEfar        & \fddedxmasserror          & \activeIerrorfar  & \activecompoundserrorfar     & \deltaerrorfar      & ---                   & \hadronicoverlapabserror  & \fdtotalerror                                          & $\delta$ param. & $I$ \\
ND&\referenceEnear       & \nddedxmasserror          & \activeIerrornear & \activecompoundserrornear     & \deltaerrornear     & \ndneutronpileuperror & \hadronicoverlapabserror  & \ndtotalerror                                         & $\delta$ param. & Pile-up \\
$\mu$C, \%&\referenceEucrel  & \muoncatcherdedxmasserror & \mucatcherIerrorrel& \mucatchercompoundserrorrel & \deltaerrormucatchrel& \muneutronpileuperror & ---                       & \mucatchertotalerror                                  & Pile-up & $\delta$ param. \\
$\mu$C, MeV\!\!\!\!&\referenceEucabs  & \muoncatcherdedxmasserrorabs\,MeV & \mucatcherIerrorabs\,MeV  & \mucatchercompoundserrorabs\,MeV  & \deltaerrormucatchabs\,MeV  & \muneutronpileuperrormev\,MeV & --- & \mucatchertotalerrorabs\,MeV & $\delta$ param. & Pile-up \\

\hline
\multicolumn{3}{l|}{Fully correlated} &                   &                           &                     &                       &                           &                               & \\

ND/FD    &\referenceEcorrnf & \sharedndfddedxmasserror  & \activeIerrornear & \activecompoundserroronegev     & \deltaerrornear     & ---                   & \hadronicoverlapcorrerror & \textbf{\sharedndfdtotalerror}& \multirow{2}{*}{$\delta$ param.} & \multirow{2}{*}{$I$} \\
$\mu$C/FD& \referenceEcorrnf      & \sharedfdmudedxmasserrormu/\sharedfdmudedxmasserrorfd  & \sharedfdmuIerrormu/\sharedfdmuIerrorfd & \sharedfdmucompoundserrormu/\sharedfdmucompoundserrorfd & \deltacorrerrormurel/\deltaerrorfar   & ---                                                 & ---                       & \textbf{\sharedfdmutotalerrormu/\sharedfdmutotalerrorfd}& \\
$\mu$C/ND& \referenceEnear & \sharedndmudedxmasserrormu/\sharedndmudedxmasserrornd  & \sharedndmuIerrormu/\sharedndmuIerrornd & \sharedndmucompoundserrormu/\sharedndmucompoundserrornd & \deltacorrerrormurel/\deltaerrornear   & \muneutronpileuperrormev\,MeV/\ndneutronpileuperror & ---                       &                          8\,MeV/\sharedndmutotalerrornd & $\delta$ param. & Pile-up \\

\hline
\multicolumn{3}{l|}{Uncorrelated}     &                   &                           &                     &                       &                           &                               & \\

ND/FD  & \referenceEuncorrnf& \nearuncorrmasserror/\faruncorrmasserror & --- & --- & --- &  \ndneutronpileuperror/--- & \nearorfarhadronicoverlaprelerror/\nearorfarhadronicoverlaprelerror & \textbf{\nearuncorrtofartotalerror/\faruncorrtoneartotalerror} & \multirow{2}{*}{Pile-up} & FD scint. \\
  && & & & & & & Ratio: \textbf{\nearfartotalerror}   & & volume \\

& & & & & & & & \\

$\mu$C/FD &\referenceEucabs & \muoncatcherdedxmasserrorabs\,MeV/---   & \mucatcherIerrorabs\,MeV/---       & ---       & \deltaerrormucatchabs\,MeV/--- & \muneutronpileuperrormev\,MeV/---  & ---/\hadronicoverlapabserror  & \textbf{\mufdtotalerrormu\,MeV/\mufdtotalerrorfd} & \multirow{2}{*}{$\delta$ param.} & \multirow{2}{*}{Steel vol.} \\
&& & & & & & & Ratio: \textbf{\mufdtotalerrormev\,MeV}            & \\

& & & & & & & & \\

$\mu$C/ND & \referenceEnear & \muuncorrtonddedxmasserror/---   & \muuncorrtondIerror/---       & ---       & \deltauncorrerrormurel/--- & --- & ---/\hadronicoverlapabserror  & \mundtotalerrormu/\mundtotalerrornd & \multirow{2}{*}{$\delta$ param.} & \multirow{2}{*}{Steel vol.} \\
&& & & & & & & Ratio: \mundtotalerror               & \\

\hline \hline
\end{tabular}
\end{center}
\end{small}
}

\newcommand{\relevancytable}{

\begin{table}
\caption{\label{tab:relevancy} Summary of how sources of uncertainty for
muon range affect other particles.  Major effects are shown in bold. }
\begin{tabular}{l c c}
\hline
\hline

Source                          & Charged hadrons     & Electrons \\

\hline
Affect true \dedx: \\
\textbf{Mass accounting}        & \textbf{Yes}        & \textbf{Yes}  \\ 
\textbf{Mean ex. energy} & \textbf{Yes}        & \textbf{Yes}  \\ 
\textbf{Density effect}         & \textbf{Yes, more so for $\pi^\pm$} & \textbf{Yes}  \\ 
Elemental composition           & Yes                 & Yes           \\ 
Compounds                       & Yes                 & Yes           \\ 
\hline
Other types of effects: \\
Hadronic modeling               & Yes (in reverse)    & Yes           \\ 
Coulomb correction              & Yes                 & Yes           \\ 
ND neutron pile-up              & Slightly            & Slightly      \\ 
{Noise modeling}                & Slightly            & Slightly      \\ 
{Muon decay}                    & {\pip and K$^+$ only} & {No}        \\ 
{Multiple scattering}           & {No}                & {No}          \\ 
{Alignment}                     & {No}                & {No}          \\ 

\hline
\hline

\end{tabular}
\end{table}

}

\newcommand{\scintmasstable}{
\begin{table}
\begin{center}
\begin{tabular}{l l c}
\hline
\hline
Error & & Near/far cancellation? \\
\hline
ASTM density & \scintdensityastm & Mostly \\
Temperature  & \densitytemperature & Somewhat \\
FD scint. volume  & \fdvolume & Partial \\
ND module volume & \ndscintvolume & Mostly \\
ND air bubbles & \ndbubbles & Partial \\
ND ``fiducial'' blends & \ndfiducialblends & No \\
\hline
FD scintillator & \fdscinterror & --- \\
ND scintillator & \ndscinterror & --- \\
ND/FD scintillator & \ndfdscinterror & --- \\
\hline
\hline
\end{tabular}
\end{center}
\caption{\label{tab:scint} Scintillator systematic uncertainties. Uses
volumetric measurements for FD, which cancel partially with ND.
}
\end{table}
}

\newcommand{\coulombcorrmuminusmevH}{1.6}
\newcommand{\coulombcorrmuminuspercentH}{0.003\%\xspace}

\newcommand{\coulombcorrmuminusmevC}{3.9}
\newcommand{\coulombcorrmuminuspercentC}{74\%\xspace}

\newcommand{\coulombcorrmuminusmevO}{4.7}
\newcommand{\coulombcorrmuminuspercentO}{3\%\xspace}

\newcommand{\coulombcorrmuminusmevCl}{7.5}
\newcommand{\coulombcorrmuminuspercentCl}{19\%\xspace}

\newcommand{\coulombcorrmuminusmevTi}{8.8}
\newcommand{\coulombcorrmuminuspercentTi}{4\%\xspace}

\newcommand{\coulombcorrmuminusmevSn}{14.8}
\newcommand{\coulombcorrmuminuspercentSn}{0.2\%\xspace}

\newcommand{\coulombcorrmupluspercentH}{20\%\xspace}

\newcommand{\coulombcorrmuplusmevC}{2.8}
\newcommand{\coulombcorrmupluspercentC}{61\%\xspace}

\newcommand{\coulombcorrmuplusmevO}{3.6}
\newcommand{\coulombcorrmupluspercentO}{3\%\xspace}

\newcommand{\coulombcorrmuplusmevCl}{6.7}
\newcommand{\coulombcorrmupluspercentCl}{14\%\xspace}

\newcommand{\coulombcorrmuplusmevTi}{8.0}
\newcommand{\coulombcorrmupluspercentTi}{3\%\xspace}

\newcommand{\coulombcorrmuplusmevSn}{14.2}
\newcommand{\coulombcorrmupluspercentSn}{0.1\%\xspace}

\newcommand{\coulombcorrmuminusmevnonothing}{4.8}

\newcommand{\coulombcorrmuplusmevnonothing}{3.0}

\newcommand{\coulombcorrmuminuserror}{0.05\%\xspace}

\newcommand{\coulombcorrmupluserror}{0.03\%\xspace}

\newcommand{\deltaerrornear}{0.7\%\xspace}

\newcommand{\deltaerrorfar}{0.8\%\xspace}

\newcommand{\steelrawmasserrormin}{0.23\%\xspace}

\newcommand{\steelrawmasserrormax}{0.32\%\xspace}

\newcommand{\steeluniformityerrormin}{0.01\%\xspace}

\newcommand{\steeluniformityerrormax}{0.03\%\xspace}

\newcommand{\steelpaintmasserror}{0.07\%\xspace}

\newcommand{\steelmasserror}{0.33\%\xspace}

\newcommand{\steelmasserrormin}{0.24\%\xspace}

\newcommand{\steelmasserrormean}{0.28\%\xspace}

\newcommand{\steelpercent}{87.6\%\xspace}

\newcommand{\pvcweighingerror}{0.46\%\xspace}

\newcommand{\plainndpvclotserror}{0.30\%\xspace}

\newcommand{\ndpvclotserror}{0.22\%\xspace}

\newcommand{\plainndpvctemporaryoffsetfudgeerror}{0.21\%\xspace}

\newcommand{\ndpvctemporaryoffsetfudgeerror}{0.16\%\xspace}

\newcommand{\glueerror}{5\%\xspace}

\newcommand{\glueerrortwoone}{25\%\xspace}

\newcommand{\scintelementerror}{0.06\%\xspace}

\newcommand{\pvcelementerror}{0.04\%\xspace}

\newcommand{\glueelementerror}{0.03\%\xspace}

\newcommand{\modeledhadronicoverlapabserror}{0.08\%\xspace}

\newcommand{\hadronicoverlapabserror}{0.19\%\xspace}

\newcommand{\hadronicoverlaprelerror}{0.05\%\xspace}

\newcommand{\nearorfarhadronicoverlaprelerror}{0.03\%\xspace}

\newcommand{\hadronicoverlapcorrerror}{0.18\%\xspace}

\newcommand{\scintdensityastm}{0.17\%\xspace}

\newcommand{\fdvolume}{0.3\%\xspace}

\newcommand{\densitytemperature}{0.1\%\xspace}

\newcommand{\ndfiducialblends}{0.09\%\xspace}

\newcommand{\ndscintvolume}{0.16\%\xspace}

\newcommand{\ndbubbles}{0.04\%\xspace}

\newcommand{\ndscinterror}{0.27\%\xspace}

\newcommand{\fdscinterror}{0.36\%\xspace}

\newcommand{\ndfdscinterror}{0.27\%\xspace}

\newcommand{\fdpvcerror}{0.46\%\xspace}

\newcommand{\ndpvcerror}{0.54\%\xspace}

\newcommand{\ndfdpvcerror}{0.31\%\xspace}

\newcommand{\ndfdscinterrorcontrib}{0.17\%\xspace}

\newcommand{\ndfdpvcerrorcontrib}{0.11\%\xspace}

\newcommand{\mufdsteelerrorcontrib}{0.29\%\xspace}

\newcommand{\mutoactivesteelerrorcontrib}{0.29\%\xspace}

\newcommand{\muoncatchermasserror}{0.29\%\xspace}

\newcommand{\fdmasserror}{0.28\%\xspace}

\newcommand{\ndmasserror}{0.26\%\xspace}

\newcommand{\nearfarmasserror}{0.21\%\xspace}

\newcommand{\nearuncorrmasserror}{0.15\%\xspace}

\newcommand{\faruncorrmasserror}{0.14\%\xspace}

\newcommand{\mufdratiomasserror}{0.38\%\xspace}

\newcommand{\mundratiomasserror}{0.37\%\xspace}

\newcommand{\muoncatcherdedxmasserror}{0.28\%\xspace}

\newcommand{\fddedxmasserror}{0.28\%\xspace}

\newcommand{\nddedxmasserror}{0.25\%\xspace}

\newcommand{\nearfardedxmasserror}{0.21\%\xspace}

\newcommand{\mufdratiodedxmasserror}{0.37\%\xspace}

\newcommand{\muuncorrtonddedxmasserror}{0.28\%\xspace}

\newcommand{\mundratiodedxmasserror}{0.35\%\xspace}

\newcommand{\sharedndfddedxmasserror}{0.23\%\xspace}

\newcommand{\sharedfdmudedxmasserrorfd}{0.28\%\xspace}

\newcommand{\sharedfdmudedxmasserrormu}{0.05\%\xspace}

\newcommand{\sharedndmudedxmasserrornd}{0.25\%\xspace}

\newcommand{\sharedndmudedxmasserrormu}{0.04\%\xspace}

\newcommand{\steeldedxpercent}{83.9\%\xspace}

\newcommand{\activeelementerror}{0.04\%\xspace}

\newcommand{\muoncatcherelementerror}{0.01\%\xspace}

\newcommand{\nominalcdedxnonothing}{1.844}

\newcommand{\shiftdncdedxnonothing}{1.853}

\newcommand{\hdedxerrI}{0.33\%\xspace}

\newcommand{\cdedxerrI}{0.46\%\xspace}

\newcommand{\odedxerrI}{0.10\%\xspace}

\newcommand{\cldedxerrI}{0.60\%\xspace}

\newcommand{\tidedxerrI}{0.12\%\xspace}

\newcommand{\sndedxerrI}{0.17\%\xspace}

\newcommand{\nominalsoupdedxnonothing}{2.072}

\newcommand{\shiftdnsoupdedxnonothing}{2.081}
\newcommand{\soupdedxerrIcorr}{0.43\%\xspace}

\newcommand{\soupdedxerrIuncorr}{0.29\%\xspace}

\newcommand{\soupdedxerrIatonegev}{0.33\%\xspace}

\newcommand{\activeIerroronegev}{0.36\%\xspace}

\newcommand{\activecompoundserroronegev}{0.20\%\xspace}

\newcommand{\activeIerrornear}{0.39\%\xspace}

\newcommand{\sharedndmuIerrornd}{0.39\%\xspace}

\newcommand{\sharedndmuIerrormu}{0.06\%\xspace}

\newcommand{\activecompoundserrornear}{0.22\%\xspace}

\newcommand{\sharedndmucompoundserrormu}{0.03\%\xspace}

\newcommand{\sharedndmucompoundserrornd}{0.22\%\xspace}

\newcommand{\activeIerrorfar}{0.37\%\xspace}

\newcommand{\sharedfdmuIerrorfd}{0.37\%\xspace}

\newcommand{\sharedfdmuIerrormu}{0.06\%\xspace}

\newcommand{\activecompoundserrorfar}{0.21\%\xspace}

\newcommand{\sharedfdmucompoundserrormu}{0.03\%\xspace}

\newcommand{\sharedfdmucompoundserrorfd}{0.21\%\xspace}

\newcommand{\mucatcherIerrorabs}{3.4}

\newcommand{\muuncorrtondIerror}{0.11\%\xspace}

\newcommand{\mucatchercompoundserrorabs}{0.6}
\newcommand{\steelIerrorrel}{0.24\%\xspace}

\newcommand{\mucatcherIerrorrel}{0.22\%\xspace}

\newcommand{\mucatchercompoundserrorrel}{0.04\%\xspace}

\newcommand{\muneutronpileuperror}{4.9\%\xspace}

\newcommand{\deltaerrormucatchrel}{1.2\%\xspace}

\newcommand{\deltauncorrerrormurel}{1.2\%\xspace}

\newcommand{\deltacorrerrormurel}{0.1\%\xspace}

\newcommand{\mucatchertotalerror}{5.0\%\xspace}

\newcommand{\muneutronpileuperrormev}{7}
\newcommand{\muoncatcherdedxmasserrorabs}{6}
\newcommand{\deltaerrormucatchabs}{18}

\newcommand{\mucatchertotalerrorabs}{21}

\newcommand{\mufdtotalerrormev}{21}
\newcommand{\mufdtotalerrorfd}{0.19\%\xspace}

\newcommand{\mufdtotalerrormu}{20}
\newcommand{\ndneutronpileuperrormev}{2}
\newcommand{\ndneutronpileuperror}{0.4\%\xspace}

\newcommand{\ndtotalerror}{0.97\%\xspace}

\newcommand{\nearuncorrtofartotalerror}{0.42\%\xspace}

\newcommand{\faruncorrtoneartotalerror}{0.15\%\xspace}

\newcommand{\nearfartotalerror}{0.45\%\xspace}

\newcommand{\mundtotalerror}{1.3\%\xspace}

\newcommand{\mundtotalerrornd}{0.19\%\xspace}

\newcommand{\mundtotalerrormu}{1.3\%\xspace}

\newcommand{\sharedndmutotalerrornd}{0.4\%\xspace}

\newcommand{\ndtotalerrorsigfig}{1.0\%\xspace}

\newcommand{\nearfartotalerrorsigfig}{0.4\%\xspace}

\newcommand{\sharedndfdtotalerrorsigfig}{0.9\%\xspace}

\newcommand{\fdtotalerror}{0.93\%\xspace}

\newcommand{\sharedndfdtotalerror}{0.91\%\xspace}

\newcommand{\sharedfdmutotalerrormu}{0.15\%\xspace}

\newcommand{\sharedfdmutotalerrorfd}{0.91\%\xspace}

\newcommand{\fdtotalerrorsigfig}{0.9\%\xspace}

\newcommand{\referenceEnear}{0.4}
\newcommand{\referenceEfar}{0.7}
\newcommand{\referenceEucabs}{2.0}
\newcommand{\referenceEucrel}{0.15}
\newcommand{\referenceEuncorrnf}{0.4}
\newcommand{\referenceEcorrnf}{0.7}

\newcommand{\genie}{\textsc{Genie}\xspace}
\newcommand{\geant}{\textsc{Geant4}\xspace}

\title{~\\
\nova muon energy scale systematic}

\author[1]{M. Strait\footnote{Corresponding author. \texttt{straitm@umn.edu}. ORCID 0000-0001-5708-8734.}}
\affil[1]{School of Physics and Astronomy, University of Minnesota Twin Cities, Minneapolis, Minnesota 55455, USA}

\author[2]{S. Bending}
\author[3]{K. Kephart}
\author[3]{P. Lukens}
\affil[2]{Physics and Astronomy Dept., University College London, Gower Street, London WC1E 6BT, UK}
\affil[3]{Fermi National Accelerator Laboratory, Batavia, Illinois 60510, USA}

\renewenvironment{abstract}
 {
  \begin{center}
  \bfseries \abstractname\vspace{0pt}
  \end{center}
  \list{}{%
    \setlength{\leftmargin}{0.75in}
    \setlength{\rightmargin}{\leftmargin}%
  }%
  \item\relax}
 {\vspace{1em}\endlist}

\begin{document}


\twocolumn[
\begin{@twocolumnfalse}
\maketitle

\begin{abstract}

The systematic uncertainty on the correspondence between muon range and energy is
developed for the \nova neutrino experiment.  \nova consists of two detectors, the Near
Detector at Fermilab and the Far Detector in northern Minnesota.  
Total errors are developed for the Near Detector,
with its Muon Catcher treated separately, the Far Detector, and all
combinations of correlated and uncorrelated errors between these three
detectors.  The absolute errors for the Near Detector
(\ndtotalerrorsigfig), the Far Detector (\fdtotalerrorsigfig), and the fully
correlated error shared by them (\sharedndfdtotalerrorsigfig) are strongly
dominated by \geant's treatment of the Bethe density effect.
At the Near Detector, the next biggest uncertainty is from stray
hits caused by neutron capture pile-up.  Other contributions are marginally
significant, with the biggest, in descending order, being due to external
measurements of the mean excitation energies of elements, detector mass accounting,
and modification of energy loss by chemical binding.  For the Muon Catcher, the
absolute error is expressed as an offset instead of a percentage: 
\mucatchertotalerrorabs\,MeV.  The density effect (at higher
energies) and neutron capture pile-up (at lower energies) are the strongly
dominant errors.  The relative error between the Near and Far Detectors is \nearfartotalerrorsigfig
and is strongly dominated by neutron capture pile-up at the Near Detector, with
a subdominant contribution from detector mass accounting.

\end{abstract}
\end{@twocolumnfalse}
]
\saythanks 

\section{Introduction}\label{sec:introduction}

In \nova analyses previous to this evaluation, the muon range uncertainties
were leading systematics for the determination of the neutrino mixing
parameters $\theta_{23}$ and $\Delta m^2_{32}$.  Often in physics analysis, large
values of systematic errors are chosen to be sure of covering true parameter
values.  However, when an error has a significant impact on an analysis, it is
appropriate to perform a careful study to reduce it.  Throughout this note, an
effort is made to enumerate and characterize the underlying uncertainties
sufficiently that no additional ``uncertainty on the uncertainty'' is needed to
account for additional undiscovered effects.  In this spirit, when the information
available is partially contradictory, an honest evaluation of the true uncertainty is made
rather than picking the largest available error.
With this evaluation, the relative
(Far/Near) muon energy scale has been reduced to a negligible error while the
absolute scale in each detector is now subdominant to other leading uncertainties.

Two quantities of interest are the absolute errors on the muon energy for the Near Detector (ND)
and Far Detector (FD) for analyses that use only one detector or the other.
The NOvA $\nu_\mu$ disappearance analysis, which uses both detectors, directly uses the component of the
error which is shared between the ND and FD, as well as the components that are uncorrelated.
The ND has two sections, one of which is nearly identical to the FD.  The second
is the Muon Catcher, which has steel planes interspersed between planes of scintillator.
All combinations of errors between the main ND, the Muon Catcher, and the FD are potentially
useful.

\sects{sec:mass} and \ref{sec:elements} work through the systematic error from knowledge of the detector mass and composition, respectively.
\sects{sec:excitation} and \ref{sec:compounds} explain what systematic errors are incurred from uncertainties in external measurements of stopping power.
\sect{density} deals with the uncertainty from the way the density effect is parameterized. 
\sect{coulomb} covers our handling of the Coulomb correction. 
\sect{hadronicoverlap} goes over hadronic response effects that can distort the track length, while
\sect{neutronpileup} explores the effects of neutron pile-up in the Near Detector. 
\sect{dontmatter} discusses some other possible concerns which turn out to be negligible in NOvA. 

\paragraph{Note on significant figures} All errors that result from a
combination of other numbers are always quoted to two digits --- even though
they may be resting on rough estimates --- so that the reader can use them in
computations of several steps without losing precision.  For the purposes of
carrying computations through this note, all digits are retained internally at
every step.  Nevertheless, the second digit of systematic errors should not
generally be taken very seriously.

\section{Mass accounting}\label{sec:mass}

The most basic uncertainty is on the amount of material a muon must traverse to
get from one cell to the next in each detector.  As the reconstruction code ultimately measures track
length in terms of cell hits, an uncertainty in the material budget translates
proportionally into an uncertainty on the reconstructed energy.  The relevant
detector components are scintillator, plastic, glue, and, for the ND Muon
Catcher only, steel.  The uncertainties on each are developed in this section
and summarized in \tab{mass}.  

\subsection{Scintillator}

The composition and production of NOvA scintillator are described in \rf{scintpaper}.

\scintmasstable

\subsubsection{Far Detector}

In the Monte Carlo model, the scintillator represents the mass of not just
the scintillator in the cells, but also the fibers that are looped through each
scintillator cell, the plastic rings that hold the fibers in place at 
one end, and the unintentional air bubbles present in many cells.
For the physical horizontal modules, there is also scintillator in the manifolds that guide the fibers from
the cells to the APDs.  This scintillator is \emph{not} modeled and must 
be subtracted from the records made during detector construction.

\paragraph{Volume} Scintillator fill was metered by volume.
Accounting for waste, the volume metered is 2\,674\,041 gallons.
The horizontal fiber manifolds contain $1.24(10)$~gallons each.  The
detectors are leak-free at a level of $10^{-8}$ over the lifetime of
the experiment~\cite{tdr}.

The scintillator density was measured by the company that did the
blending at $60^\circ$F.  Using tables that give the density of mineral
oil as a function of temperature, this is corrected for the actual
mean temperature of each detector.  There were 25 separate scintillator batches, 
each with a density measurement.  For each FD module, the batch
or batches used to fill it were tracked, and using this the average density was determined to be 
0.8529\,g/cm$^3$.  Taking into account that the Monte Carlo scintillator
includes the denser fibers, the modeled density is the slightly higher
0.8530\,g/cm$^3$.  The ND filling was not tracked as closely, but only
a few batches of scintillator were used.  The effective
density of ND scintillator is 0.8576\,g/cm$^3$.

The FD horizontal modules could not be filled completely because, as 
extruded plastic, they are not perfectly straight and in many instances
air remained trapped at the top.  From the filling data, 
the horizontal modules are estimated to be 98.75(30)\% full of scintillator.  Because
of limitations of the NOvA software, the modules in the Monte Carlo are represented as being 
completely full of scintillator at the density given above.  Corrections
are then applied at the analysis level to account for the reduced mass.

The RMS of the metered volume for FD vertical modules, which can be reliably filled completely, is 0.78\%.  The majority of this variation is assumed to be
real differences between modules and/or ``statistical''
error that tends to cancel out when adding together many modules.  The fourteen flow meters
used were calibrated by Great Plains Industries.  Using the quoted errors
and the spread of these errors, the error is estimated to be
\fdvolume.   This is assumed to be the dominant source of error for scintillator
volume in the horizontal modules.

For the fully-filled vertical modules, besides the flow meter data, 
volumetric and linear measurements of the plastic extrusions are available.  The data
therefore overconstrains the problem.  These three sources of information
agree at the $\sim1.5\sigma$ level.  The volumetric measurements are chosen
because they have the smallest stated uncertainties, as well as because they
give a result between the other two. The error estimated from the spread of
several measurements is $\pm 0.16\%$, which we increase to 0.3\% given the
slight tension between methods.

\paragraph{Density} The error from the manufacturer on the densities of
each scintillator batch is not well defined, but from
the distribution of the measurements, it is taken to be
\scintdensityastm.  There is an additional error on the density, and so on the relevant mass in the
detector, from temperature variations.  Each $^\circ$F decreases the density by 0.04\%,
whereas the density of PVC changes much less: 0.008\%~\cite{etwebnonote}.  Given a
few degrees uncertainty in the detector temperature, this incurs a systematic
of \densitytemperature.

\subsubsection{Near Detector}

The shorter ND horizontal modules are probably completely filled, because shorter sections
of extrusion are likely to be straight enough to avoid trapping air.
The possibility that there is some air in them is accounted for
by taking them to have a uniform probability of 0--20\% as much as the
FD per unit length. This is arbitrary, but as it has a negligible effect on the total error (see
\tab{scint}), it does not merit further study.  The ND shares the FD's systematic errors on the density, but has
a different error on the volume which comes from the extrusion volume instead
of the metering since only the FD scintillator was metered.

Three of the 25 blends of scintillator were used in the ND.  While it is known
how much of each blend is in the ND, it isn't clear where in the detector each
of the three are.  Since event selection effects in the ND can be quite strong,
these blends could have effective importance weights which are significantly
different from their raw fractions.  Assuming the worst case that only an
unknown one of the three is in the effective fiducial volume gives an error of
0.18\% (the RMS of the appropriate three-point distribution) on the effective
scintillator mass.  Half of this is taken as a reasonable figure for any
realistic selection, since selected events include muons that traverse
nearly all parts of the detector.

The uncertainties on scintillator mass are summarized in \tab{scint}, which
also gives the total errors on scintillator mass in each detector and the
error on the ratio of scintillator masses.

\subsection{PVC}

\begin{table}
\begin{center}
\begin{tabular}{l l c}
\hline
\hline
Error & & Near/Far cancellation? \\
\hline
Scales & \pvcweighingerror  & Mostly \\
ND lots  & \ndpvclotserror  & No \\
ND apparent offset  & \ndpvctemporaryoffsetfudgeerror  & No \\
\hline
FD & \fdpvcerror & --- \\
ND & \ndpvcerror & --- \\
ND/FD & \ndfdpvcerror & --- \\
\hline
\hline
\end{tabular}
\end{center}
\caption{\label{tab:pvc} PVC mass systematic uncertainties.}
\end{table}

\subsubsection{Far Detector}

The PVC extrusions were weighed by the extruder, Extrutech, using a Dillon
EDxtreme Digital Dynamometer.  The masses are stored in the NOvA hardware
database.  An \mbox{EDx-1T}, which has a ``standard resolution'' (i.e. readout
granularity) of 2 pounds was used initially, and then later an \mbox{EDx-5T}
with a ``standard resolution'' of 5 pounds.  Both of these models have an
``full scale accuracy'' of 0.1\% and have NIST-traceable
calibrations~\cite{dillonnonote}.

This must be translated into $1\sigma$ errors for the weights of
NOvA extrusions.  Since full scale is either 2200 pounds or 11\,000 pounds, the
``accuracies'' are 0.43\% and 2.1\% at 515 pounds, where \emph{accuracy} 
means that these scales are calibrated so that they are guaranteed to fall within the
stated range.  Assuming a uniform distribution within the range, the RMS values are
0.25\% and 1.2\%.  As 8200 of 23\,400 extrusions were weighed with the less
accurate scale, the overall uncertainty is \pvcweighingerror, assuming the calibrations
were uncorrelated.  (If instead full correlation is assumed, 0.59\%, but we 
see no reason to assume this.)

At one point along the extrusion process, Argonne independently weighed some
extrusions and obtained an average 1 pound different
than Extrutech (0.2\%).  This confirms that the above error is reasonable.

Because the readout steps are not too small compared to the variation in
extrusion weights, it is possible that additional error could be incurred from rounding,
even if the measurements were unbiased, especially if the distribution of
masses weren't Gaussian.  However, it is quite Gaussian. A toy Monte Carlo
confirmed that this effect is totally negligible.

Another conceivably relevant effect is the ambiguity between weight and mass.
The scale is presumably calibrated against standard gravity,
9.806\,65\,m/s$^2$.  The gravity at Extrutech in Manitowoc, WI is 0.02\% less
than standard~\cite{gravity1nonote,gravity2}, so the difference is negligible.

\begin{figure}
\begin{center}
\includegraphics[width=\columnwidth]{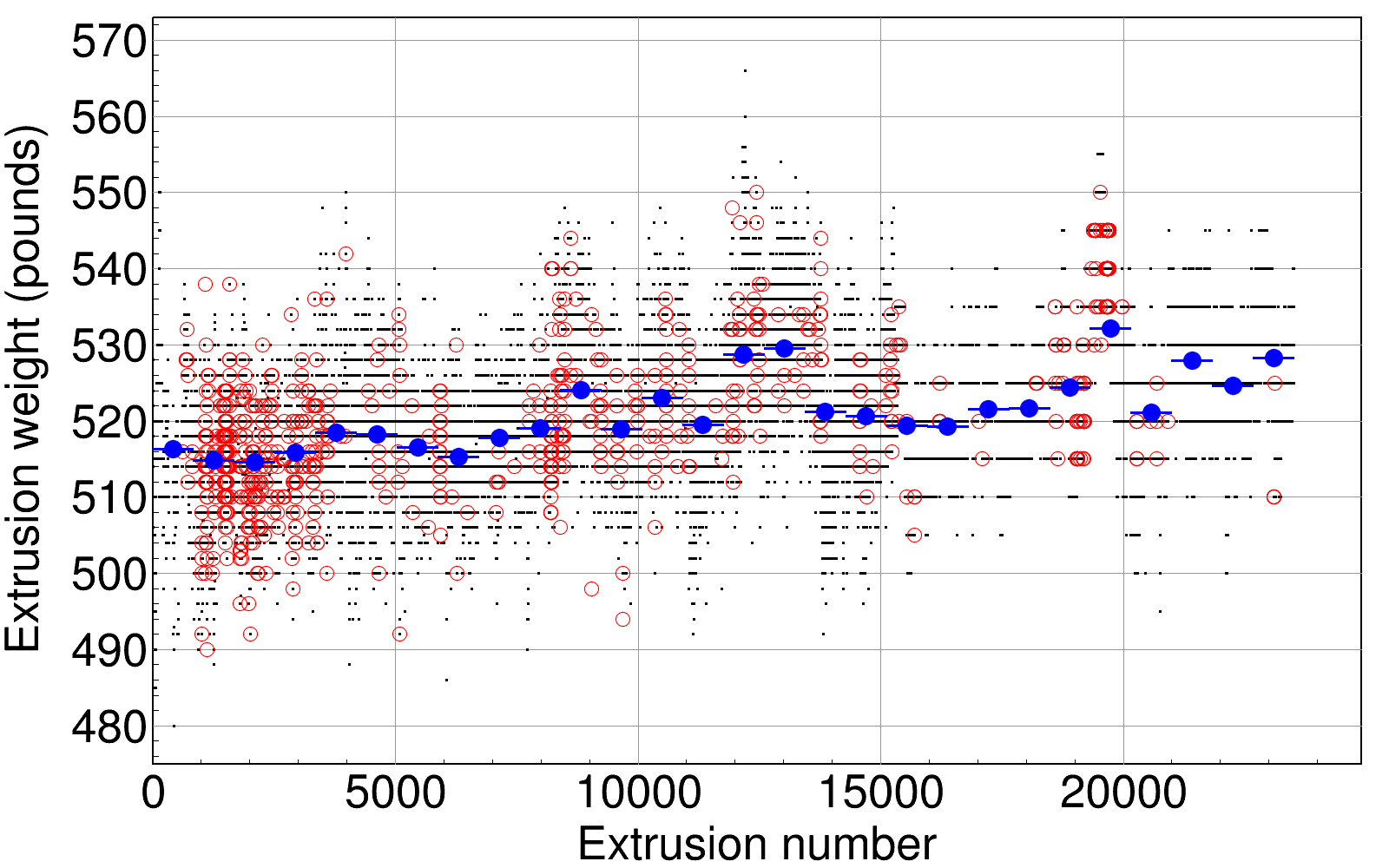}
\end{center}
\caption{\label{fig:extrusionweight}
Extrusion weights (black dots), with the earliest extrusions produced on the left.  The open red
circles are those marked ``Near Detector'' or ``scrap'', i.e.  those eligible
for use in the ND.  The filled blue circles are averages for each 28th part of the
production, meant to roughly correspond to each FD block.  }

\end{figure}

\begin{figure}
\begin{center}
\includegraphics[width=\columnwidth]{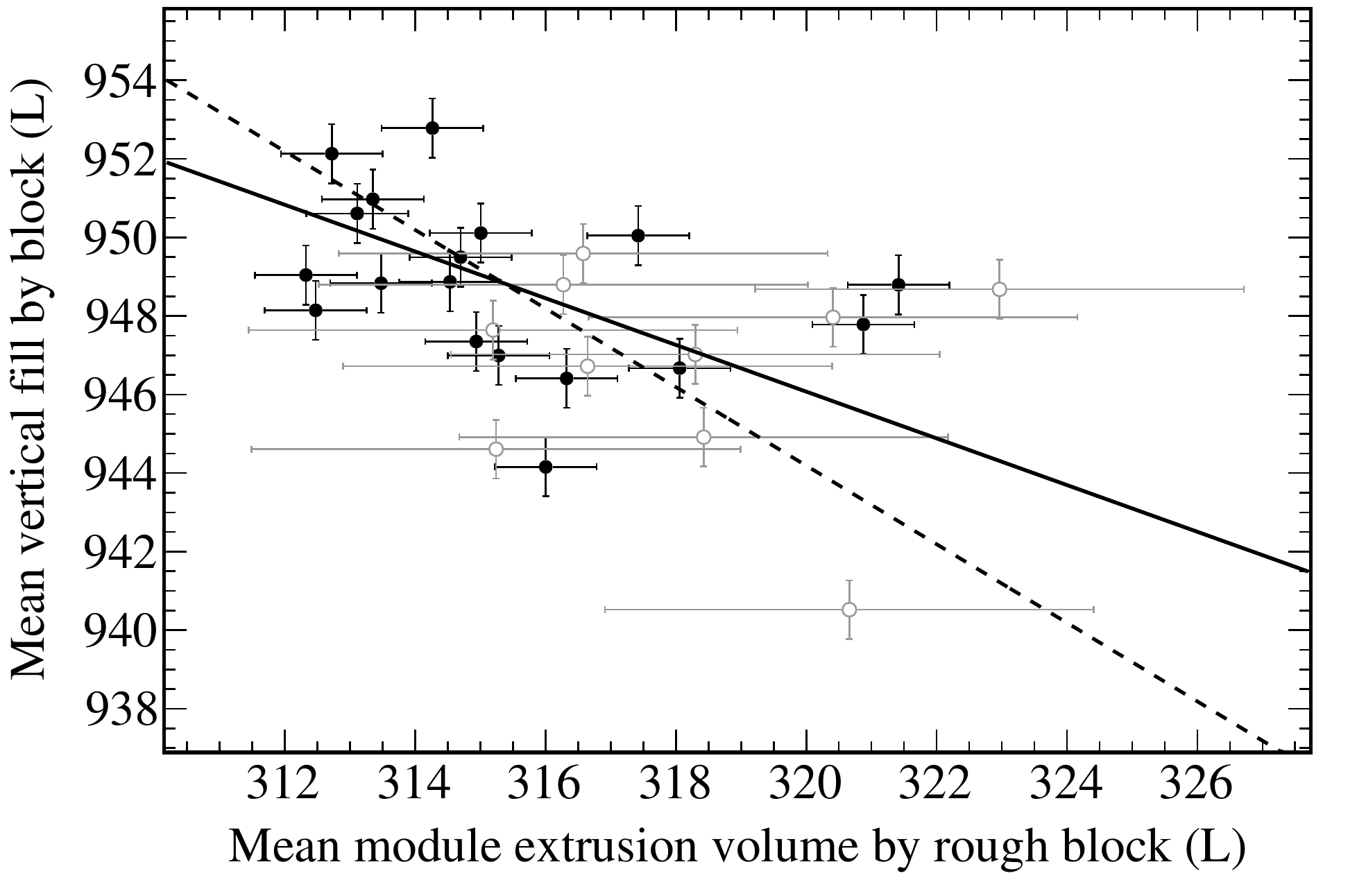}
\end{center}
\caption{\label{fig:volumeweight} \protect\input{mufig/volumeweight.txt}
}
\end{figure}

\subsubsection{Near Detector}

While NOvA's hardware database stores exactly which extrusions were used in the FD, the much shorter ND
extrusions were obtained by using scraps of FD-length extrusions.  The database identifies
which extrusions were marked as being usable for the ND, but not how many
ND modules were made from each. If the ND extrusions were drawn uniformly from
the eligible full-length extrusions, they would have a mass per unit length
\plainndpvctemporaryoffsetfudgeerror lower than the mean FD extrusion (extrusion mass tended to increase over
time, see \fig{extrusionweight}), but it is not clear that this is the correct
description.

Each full-length extrusion could have contributed between zero and three
ND-length extrusions to the final detector.  Suppose that this number tended to
increase from around zero to around three over the course of production.  This
would raise the estimate of the PVC mass by 0.5\%.  Taking a uniform distribution
between this trend and the reverse gives an RMS error of \plainndpvclotserror.

It is not clear why the extrusions increased in mass over the course
of production, but if it is because the walls tended to be thicker, then there
would be less internal volume for scintillator, partially cancelling the
increase in stopping power from the additional plastic.  Scintillator was not metered for the ND fill, but
the FD filling data can provide some information. 
As shown in \fig{volumeweight}, there is probably an
anti-correlation between extrusion mass and scintillator fill volume for FD
vertical modules.  There are strong correlations between the data points, so it
is not completely clear if this is a real effect, but it is clear that there is
no \emph{positive} correlation, at least.  Letting there be a 50\% probability of no
correlation and 50\% probability that each additional liter of PVC means one
fewer liter of scintillator, the effective error from not knowing
exactly what set of extrusions were used is reduced to \ndpvclotserror.
The effective offset in the case of uniformly drawing from all 
eligible extrusions is reduced to \ndpvctemporaryoffsetfudgeerror.
These are added in quadrature when finding the total \dedx error.

Uncertainties on PVC mass are summarized in \tab{pvc}.
In the tables, all of this error is attributed
to PVC, even though it has something to do with the scintillator as well.

\dedxtable

\massaccountingtable

\subsection{Glue}

The glue mass that holds two extrusions together into a module is estimated from
measurements of the glue cross section in completed modules.  This is about 4\% of
the total glue and is assigned a \glueerrortwoone error.  This is negligible.

The other 96\% of the glue is that which holds each plane to the next.
Measurements taken during construction 
give a mass of 9.05\,kg per plane.  There is some uncertainty due to the mass
lost from outgassing, estimated at \glueerror.

\subsection{Muon Catcher}

The ND Muon Catcher is made of alternating layers of steel planes and two
active (scintillator) planes.  The mass of the steel in the fiducial region of the Muon Catcher was carefully
measured~\cite{novadoc16937}.  \steeltext

The error on the mass of the active planes in the Muon Catcher is the same as
that for the main ND.  The steel planes are \steelpercent the mass of the Muon
Catcher, and \steeldedxpercent of the stopping power at minimum ionization.
The error in the steel mass is uncorrelated to the \nearfardedxmasserror error
in the active plane mass.  The combined error for the steel planes and active
planes is \muoncatcherdedxmasserror.

\subsection{Summary}

\tab{mass} summarizes the mass accounting errors.  As the quantity of interest 
is the mean \dedx integrated through a large number of detector planes, the total errors
on \dedx that stem from mass accounting are weighted by the nominal  \dedx of each material.  These
values are shown in \tab{dedx}.  Of note is that the \dedx of steel is much lower
than of hydrocarbons, because of the lack of hydrogen, so when computing the weighted
averages, the impact on the error from the mass of steel is somewhat reduced.

\section{Elemental composition}\label{sec:elements}

How much can \dedx of each component of the detector change via reasonable
shifts in elemental composition?  The hydrogen fraction has the biggest impact
since hydrogen has a factor of two greater stopping power than the other
components.

\subsection{Active planes}

\subsubsection{Scintillator}

Renkert 70T mineral oil is by far the largest
component of the scintillator. Its hydrogen-to-carbon ratio is nominally 2.056. 
All of the molecular components have a similar ratio.
Rather large adjustments to the fractions of each component 
within reasonable bounds can move this by about 0.01, which shifts
the \dedx of the scintillator by \scintelementerror.

Other scintillator components have well defined chemical formulas and cannot be
adjusted except to change their relative proportions.  None of these has any
significant effect.  Since the oil and the
pseudocumeme are both pure hydrocarbons, adjusting their relative fractions
makes very little difference.  The third biggest component, PPO, has oxygen and
nitrogen, but it is only $\sim0.1\%$ of the scintillator and the amount is well
known in any case.

\subsubsection{Plastic}

The PVC-based plastic has only one component with an uncertain
elemental make-up, which is paraffin wax.  This is similar to mineral oil,
except with a longer average chain length.  As it is less than 1\% of the plastic
mass, the error incurred by uncertainty on the chain lengths is negligible.

The second biggest component of the plastic after PVC itself is titanium dioxide.
This does not have hydrogen, whereas all of the other components do, so it
must be examined carefully.
It is nominally 19 parts
out of a 129 part blend~\cite{TALAGA201777}.  It is reasonable to assume that this is accurate to
at least half a part.  Letting it be 18.5 or 19.5
changes the \dedx of the plastic by 0.03\%.

Similarly, one can consider adjusting the PVC content of the plastic between 95 and
105 parts, giving a shift of only 0.03\% in the \dedx for even shifts of
such an unlikely size.  In all, \pvcelementerror is a conservative estimate of the error
from the elemental composition of the plastic.

\subsubsection{Glue}

The exact composition of the glue is a trade secret; ranges are given for the
major components.  No realistic  variations in the composition change the glue
\dedx by more than \mbox{$\sim\glueelementerror$}.  Given the small mass of
glue, this is quite negligible. 

\subsubsection{Total}

Putting scintillator, plastic and glue together, a rather conservative error
for the \dedx due to elemental composition for the FD and the main ND is
\activeelementerror.

\subsection{Steel, Muon Catcher}

Since the steel is a commodity, ASTM A36 steel, its composition is very well
known.  Further, without any hydrogen to worry about, changes in which metals are
present have very little effect.  At most the uncertainty from elemental
composition on \dedx could be 0.01\%.  Since the steel contributes a large
majority to the Muon Catcher stopping power, the overall Muon Catcher error
from elemental composition is also, at most, \muoncatcherelementerror.

\section{Elements' mean excitation energy}\label{sec:excitation}

The 2016 Review of Particle Physics by the Particle Data Group (PDG) section 33.2.3 says that ``the `Bethe equation'\,''
(33.5) ``describes the mean rate of energy loss in the region $0.1 \lesssim
\beta\gamma \lesssim 1000$ [0.5\,MeV--100\,GeV for muons] for intermediate-Z
materials with an accuracy of a few percent''~\cite{pdg2016}.  

This sentence requires some unpacking.  We interpret ``in\-ter\-me\-di\-ate-Z'' to mean that
for hydrogen and heavy elements, the energy range in which the bare Bethe
equation is a good approximation are different. Our Monte Carlo does not use
the bare Bethe equation in any case, but also includes further low and high
energy corrections, so the presence of hydrogen in NOvA is not a problem.
``Accuracy of a few percent'' means, first, that the Bethe equation does not
apply exactly at low energies where shell effects are important, nor at high energies
where radiative effects are important.  Neither of these effects exceed a few
percent within the given energy range.  \geant \emph{does} model shell and
radiative effects~\cite{geantmanual}, so the lack of them in this formula does
not represent an inaccuracy in NOvA's Monte Carlo.  Each of them are, in any
case, a small contribution in NOvA's energy range.  Second, it means that the
equation has been tested to an accuracy of a few percent.

As far as the testing of the theory is concerned, a study, which specifically
deals with the individual collisions that the Bethe formula is based on,
says~\cite[p. 189]{Bichsel:2006cs}, ``I suggest that the calculations presented
[\dots] agree with measurements to about 1\%.  Thus the results of the theory
described here should be taken seriously. If differences between measurements
and calculations appear, explanations should be found.''  In other words,
the equation itself represents a theory which has passed experimental tests
and should be taken as correct within its range of validity. Any
inaccuracies arise from the values of the \emph{terms} of the formula.

According to PDG 2016, Sec.~33.2.4: ``\,`The determination of the mean
excitation energy [$I$] is the principle non-trivial task in the evaluation of
the Bethe stopping-power formula'\,''.  The term for mean excitation energy for
individual elements is covered in this section, followed by \sect{compounds}
which discusses complications for compounds, and \sect{density} which deals
with the density effect term.

The parameter $I$ enters into the Bethe formula as $\log(1/I^2)$.
For values of $I$, the PDG cites ICRU report 37~\cite{icru37}. 
For instance, carbon's value is given as $I = (78 \pm 7)$\,eV.  But the uncertainty is not
$1\sigma$.  A footnote on page 19 explains that it is:

\begin{displayquote}

\dots arrived at by
subjective judgments, and with a meaning that is not easily defined.
One possible interpretation would be the following.  If, in the future,
the measurement accuracy were to be sufficiently improved so that $I$-values
could be determined with an order of magnitude better than at present, we
expect that for perhaps 90\% of the cases \dots the future $I$-values
would lie within the limits of uncertainty given in this report. 
The reader \dots could convert them to ``standard deviations'' by
multiplying that by a factor of about one half.

\end{displayquote}

This is a bit inconsistent since 90\% is only 1.64$\sigma$, not 2$\sigma$.  The difference,
presumably, was seen as immaterial given the sort of estimations needed to summarize
several 1950's-style experiments with poorly constrained
systematic errors.  To be conservative, we divide by
1.64 rather than 2.  For instance, the $1\sigma$ error used for
carbon is $\pm 4.3$\,eV.

\subsection{Active planes}

\begin{figure}\begin{center}
\includegraphics[width=\columnwidth]{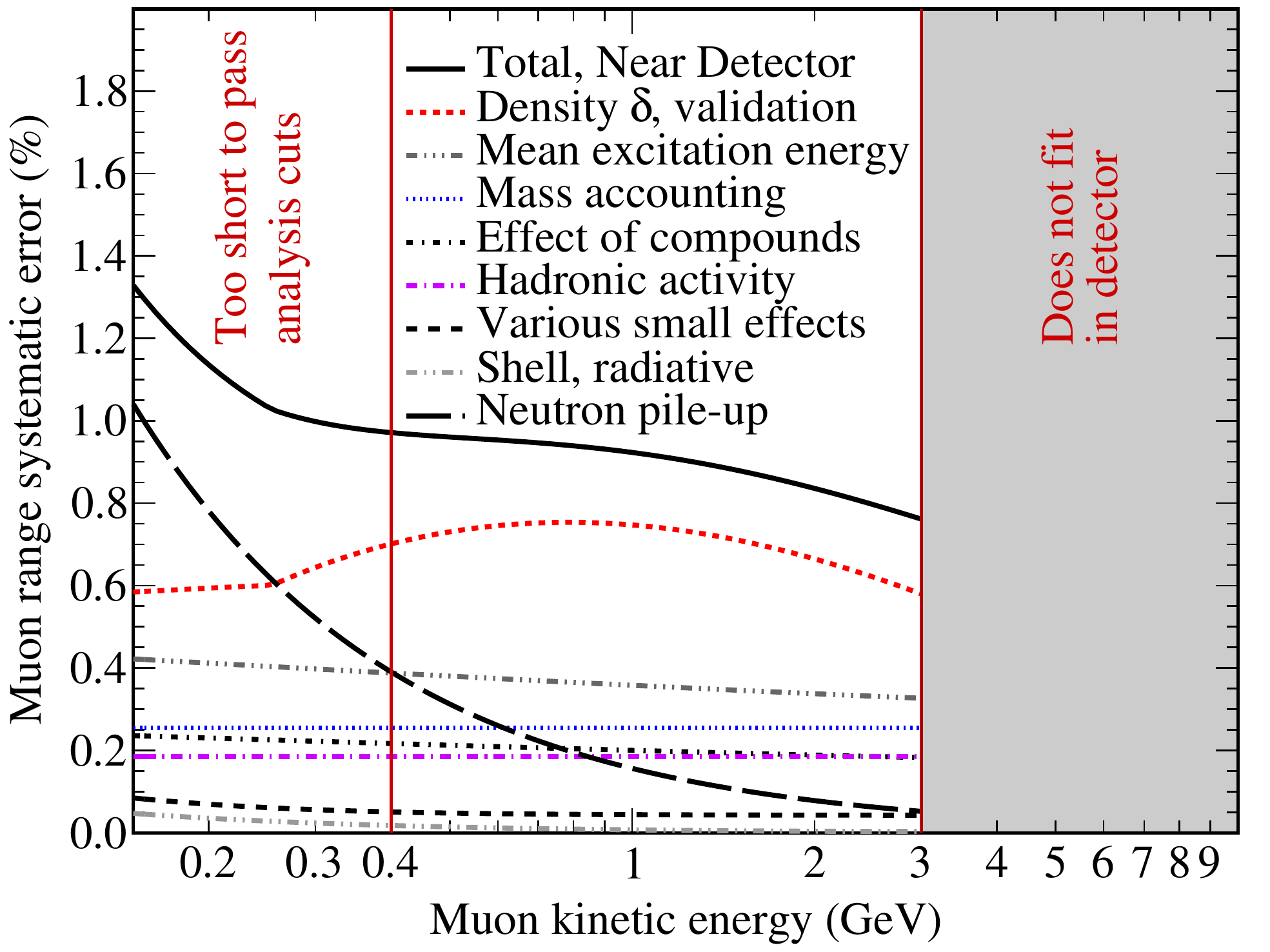}
\end{center}
\caption{\label{fig:ndplot} \protect\input{mufig/ndplot.txt}}
\end{figure}

Let us consider the effect of $I$ value uncertainties on muon range at 1\,GeV,
roughly the average muon energy from NuMI interactions.  As shown in
\fig{ndplot}ff, the resulting uncertainty on muon range is nearly constant over
the range of NOvA energies.  The nominal \dedx for carbon (graphite, the only
form of carbon with experimental data) at 1\,GeV is
\nominalcdedxnonothing\,MeV\,cm$^2$/g.  If $I$ is reduced by its error, the
\dedx increases to \shiftdncdedxnonothing\,MeV\,cm$^2$/g. Therefore the error
from $I$ is \cdedxerrI. 

Similarly for hydrogen (liquid) it is \hdedxerrI and for
chlorine (liquid) it is \cldedxerrI. For chlorine the tables quotes only
174\,eV with no error since there is no experimental data.  We use 12\,eV 
as an estimate for the
error on chlorine, which is double the largest nearby error in the
table.  Titanium and oxygen both have data on the value of $I$ and have \dedx
errors from it at 1\,GeV equal to \tidedxerrI and \odedxerrI, respectively.
For tin --- a very minor component of the detector at 0.12\% of the mass, but mentioned because
it is so much heavier than anything else --- the error is \sndedxerrI.

Neglecting other elements for this purpose, the nominal \dedx of the main detector
at 1\,GeV is \nominalsoupdedxnonothing\,MeV\,cm$^2$/g.  If all $I$ values
are taken to be fully correlated and are shifted down by $1\sigma$, 
it is instead \shiftdnsoupdedxnonothing\,MeV\,cm$^2$/g, which is 
a difference of \soupdedxerrIcorr.

But it is too conservative to consider the errors fully correlated.  The major
components are hydrogen, carbon and chlorine.  First
of all, chlorine's error is an arbitrary large value inspired by periodic trends,
so it is largely uncorrelated with hydrogen or carbon (although possibly
somewhat correlated with the nearby titanium, but titanium is only 3.2\% of
NOvA's mass, so this is negligible). 

The adopted value of hydrogen's $I$ is a combination of measurements that were
normalized to copper, those normalized to water, and an absolute
measurement~\cite{icru37}.  The absolute measurement is the most precise.  The
largest source of error on the absolute measurement, which used muon and proton
range, is from the uncertainty of the pressure of liquid hydrogen in the
Argonne 30-in bubble chamber~\cite{garbinciushyman1970}.

The adopted value of carbon's $I$ is a combination of measurements normalized
to copper, water, and aluminum as well as an absolute measurement.  The
absolute measurement is the most precise, followed by those normalized to water
and aluminum.

The absolute measurements in each case used a direct measurement of particle
range, with different energies used for the two elements.  There's no overlap
in authors between measurements used for carbon and hydrogen.  And the major
systematic error on the best hydrogen measurement cannot affect the carbon
measurement.  In summary, it is
fair to consider the errors to be mostly uncorrelated.  The error if the
elements were fully uncorrelated, would be 
\soupdedxerrIuncorr.  We have somewhat arbitrarily chosen to take a point 1/4
of the way between uncorrelated and fully correlated, arriving \soupdedxerrIatonegev
for the \dedx at 1\,GeV.

The quantity desired is the error on the conversion between track
length and initial muon energy.  For this, the error from zero to 1\,GeV,
weighted by the reciprocal of the nominal \dedx, is integrated.  The result is
slightly higher because the error rises at low energy: \activeIerroronegev.  This
integrated error is shown as a function of energy in \fig{ndplot}ff.  Despite the
energy dependence, since the overall muon range error is dominated by the density effect
parameterization (see \sect{density}), the range error is ultimately taken
as a single number evaluated at 1\,GeV.

\begin{figure}\begin{center}
\includegraphics[width=\columnwidth]{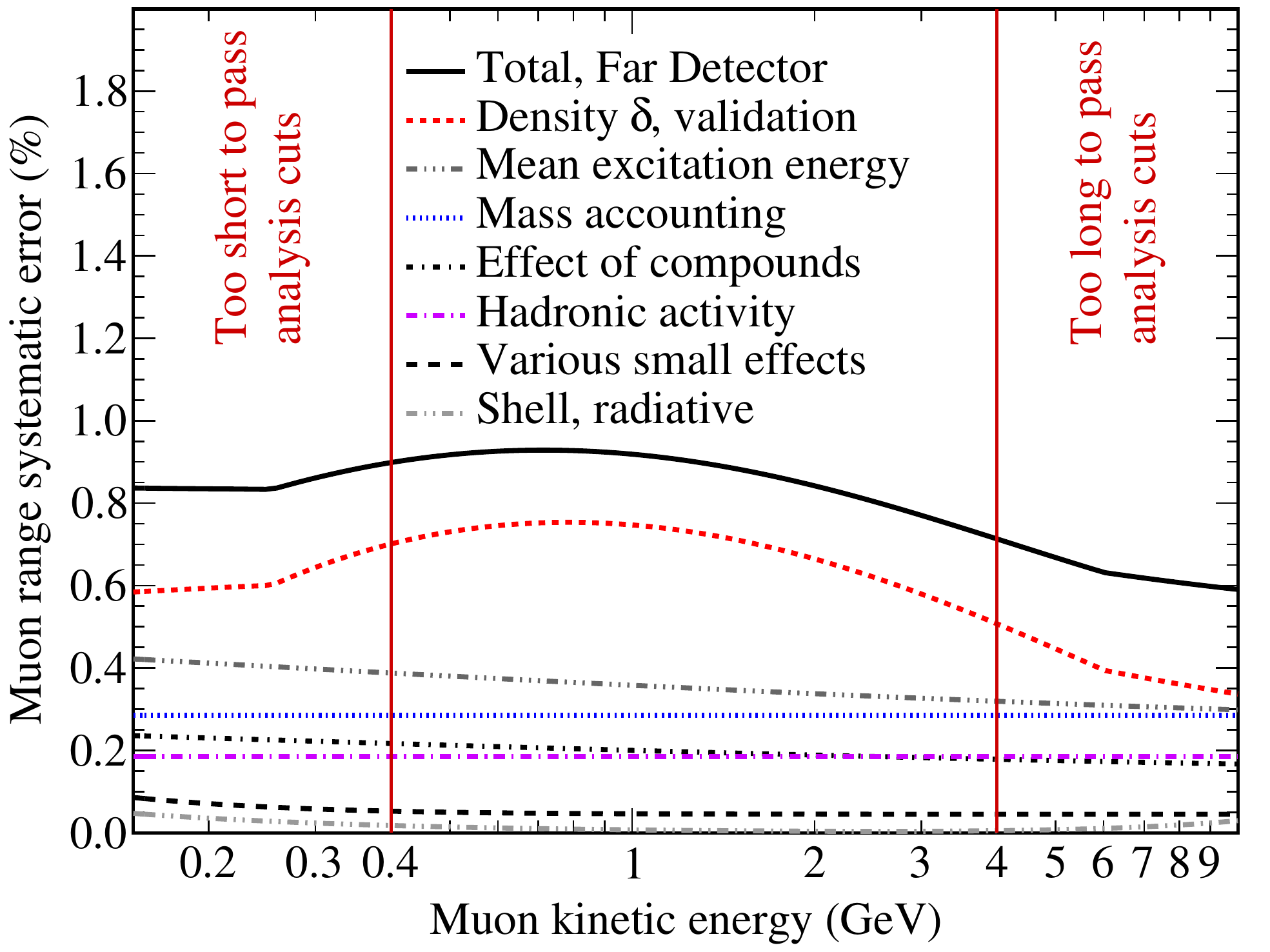}
\end{center}
\caption{\label{fig:fdplot} \protect\input{mufig/fdplot.txt}}
\end{figure}

\begin{figure}\begin{center}
\includegraphics[width=\columnwidth]{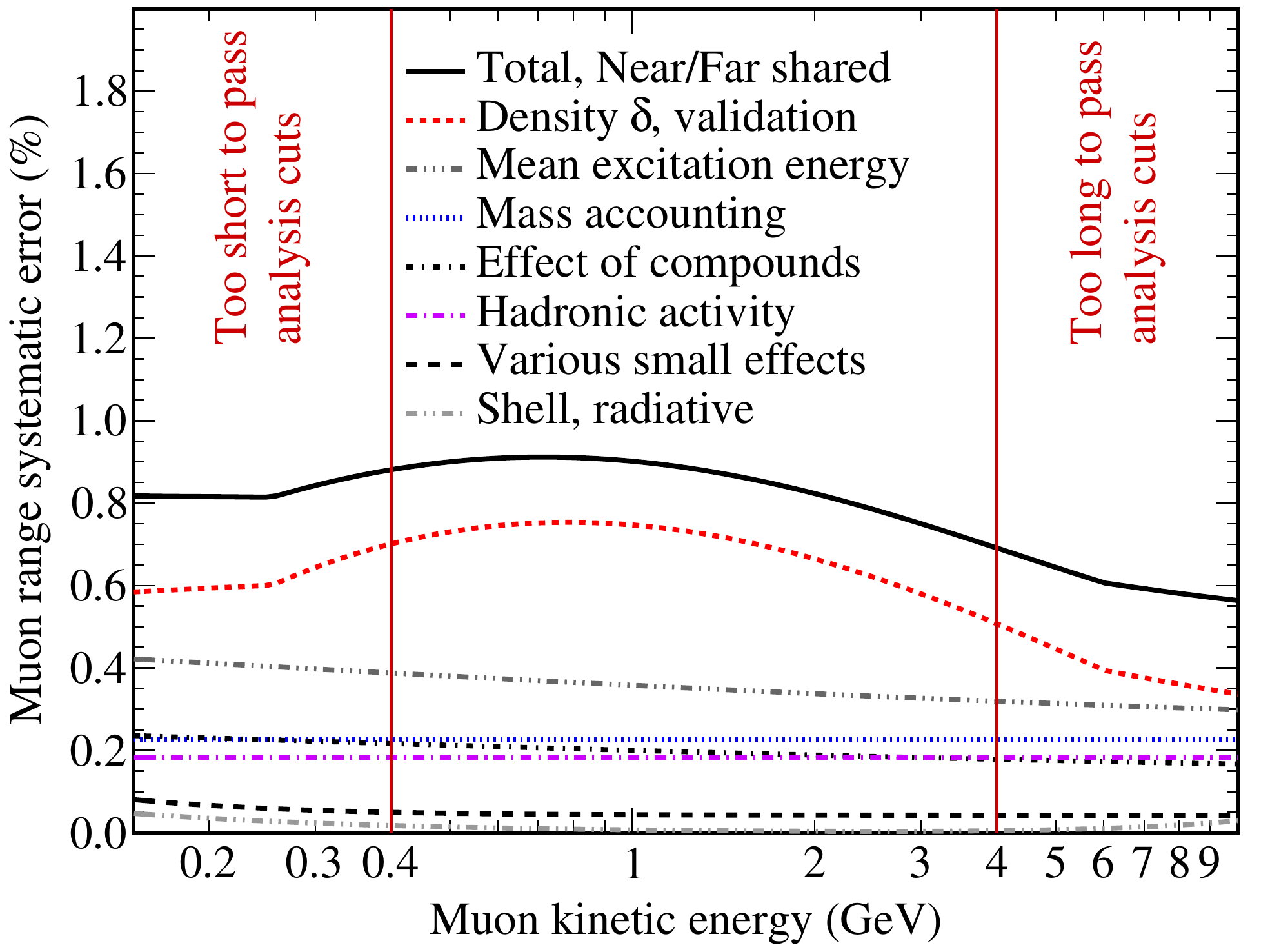}
\end{center}
\caption{\label{fig:corr_nf_plot} \protect\input{mufig/corr_nf_plot.txt}}
\end{figure}

\begin{figure}\begin{center}
\includegraphics[width=\columnwidth]{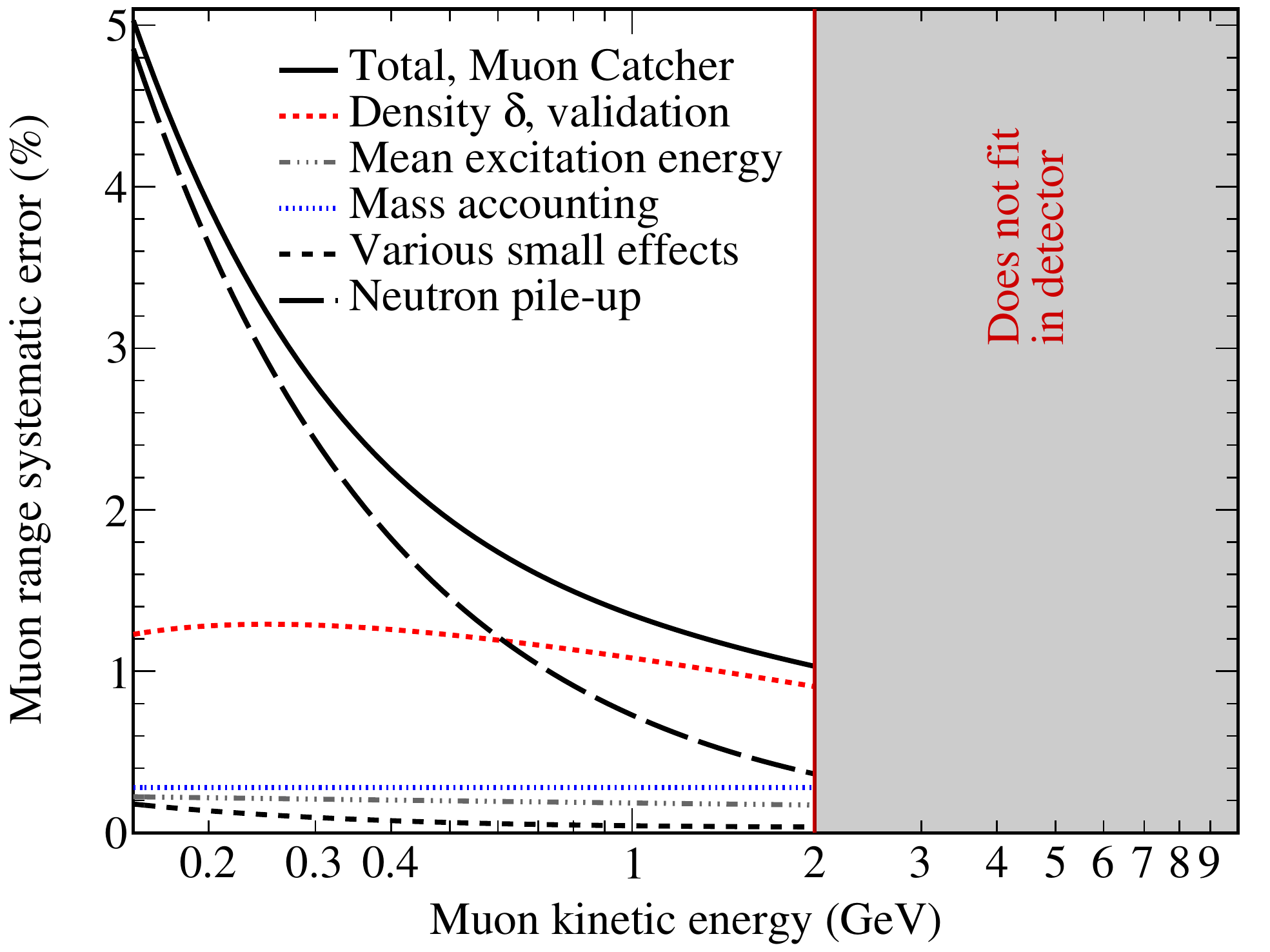}

\includegraphics[width=\columnwidth]{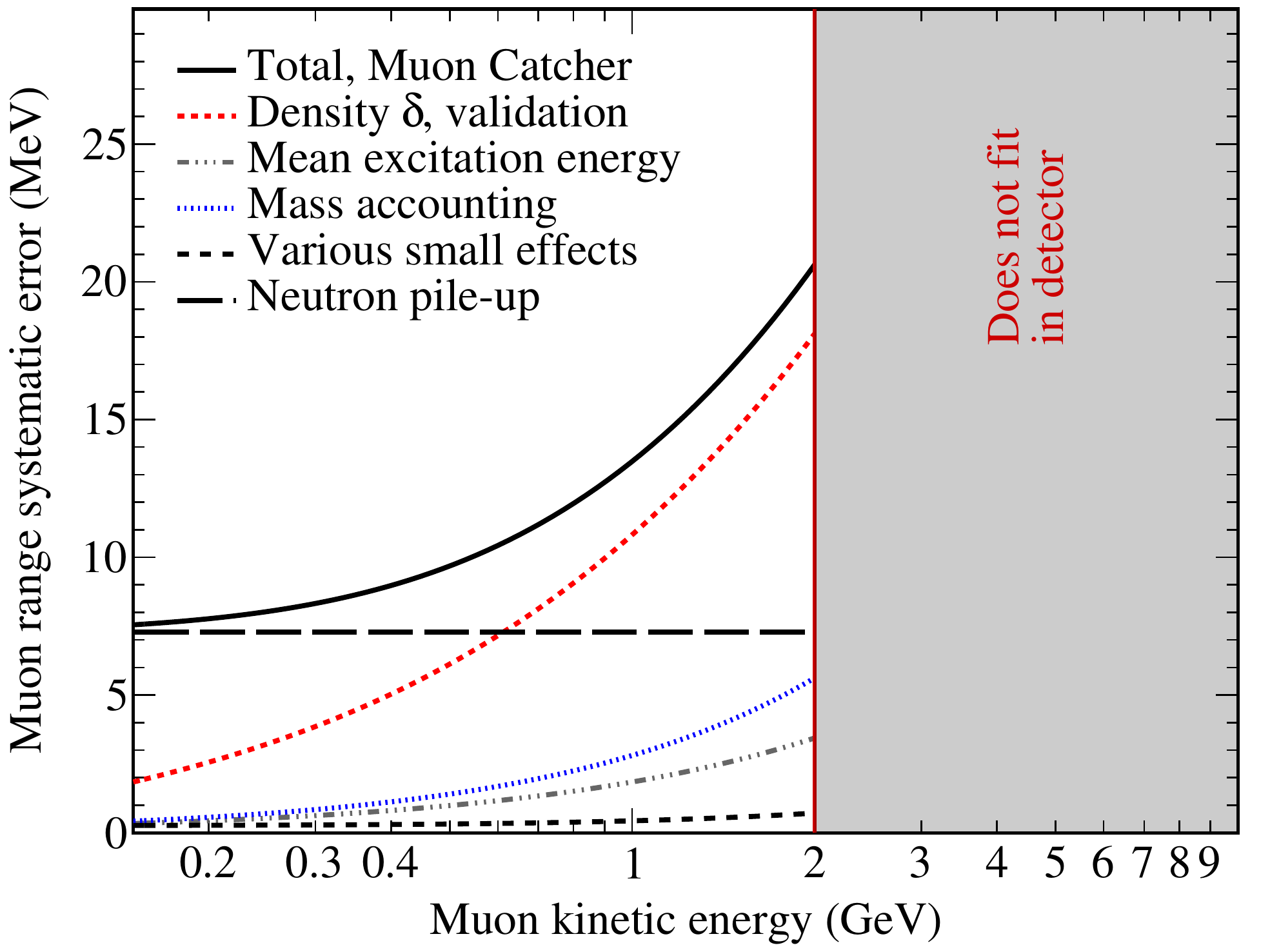}
\end{center}
\caption{\label{fig:ucplot} \protect\input{mufig/ucplot.txt}}
\end{figure}

\begin{figure}\begin{center}
\includegraphics[width=\columnwidth]{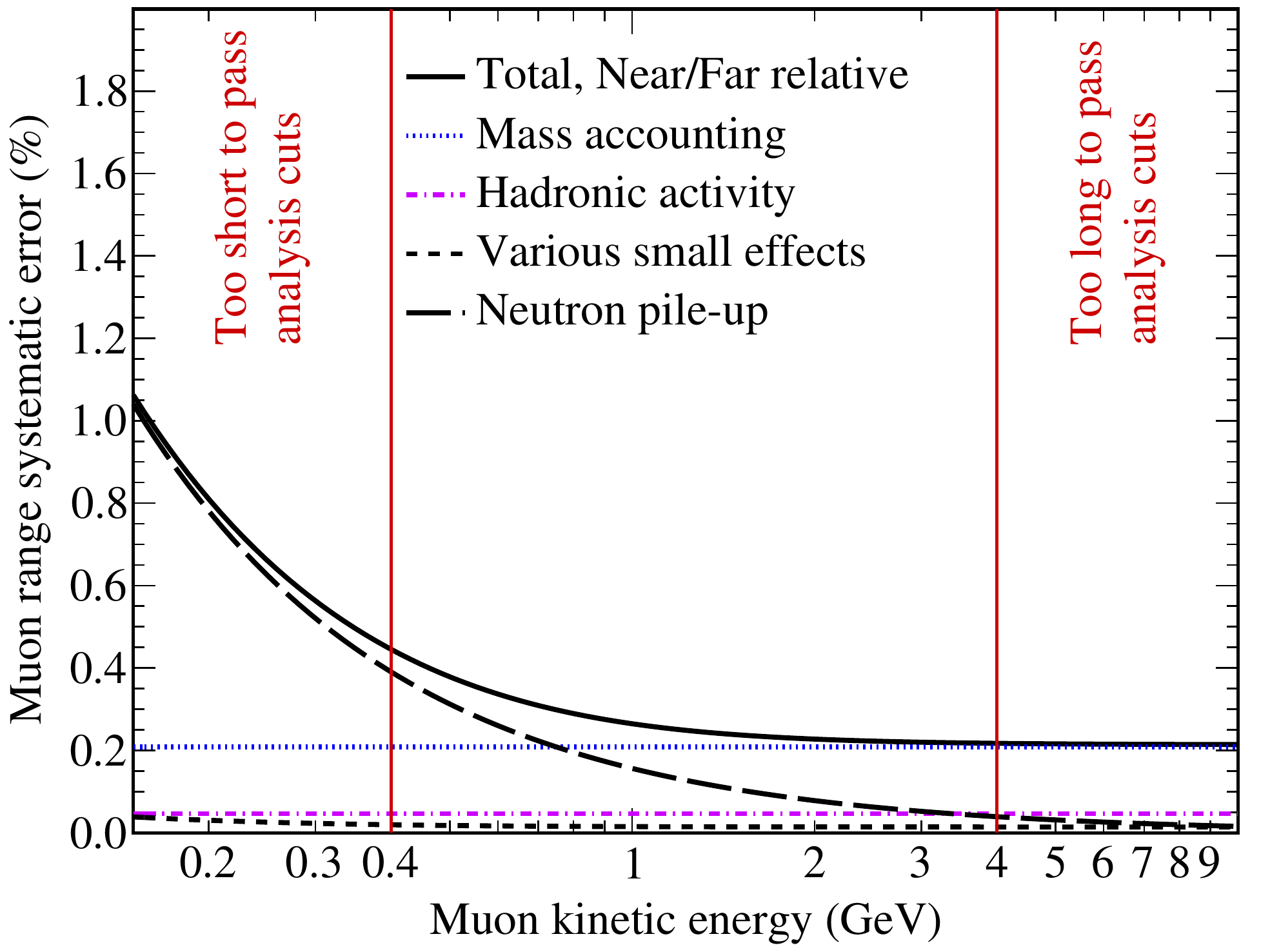}
\end{center}
\caption{\label{fig:uncorr_nf_plot} \protect\input{mufig/uncorr_nf_plot.txt}}
\end{figure}

\subsection{Steel, Muon Catcher}

The error on iron's $I$ is quite small: $(286 \pm 5)$\,eV.  Manganese is the
next largest component at 1.03\%.  Its $I$ error is also small, $(272 \pm
6)$\,eV.  Because the best measurements in each case come from the same
paper~\cite{icru37}, we assume they are fully correlated.  The error on \dedx in
the steel from $I$ values is then \steelIerrorrel. 

The measurements of $I$ for iron and manganese are mostly unrelated to those on
carbon and hydrogen.  There is one paper in common, \rf{bakkersegre1951}, but
it is the oldest and least precise.  The most precise measurements of iron
come directly from particle ranges without reference to other materials.
The steel and active planes are therefore combined assuming the errors are
mostly uncorrelated, and arrive at 
\mucatcherIerrorrel for the Muon Catcher.

\section{Compounds}\label{sec:compounds}

What about the fact that the detector is made of hydrocarbons and not a mixture
of graphite and liquid hydrogen?  PDG 2016 Sec.~33.2.10 says that it leads to
an underestimate of $I$ (overestimate of \dedx) if the compounds are treated as
mixtures.  They reference \rfs{SELTZER19821189,SELTZER1984665} by Seltzer and
Berger, from which it would seem that the mistake can go either way.

NOvA's \geant model does not specify how elements are bound together into
compounds.  The manual~\cite{geantmanual}, Sec. 12.2.2: ``Charged Hadron
Incident: Low energy extensions: Energy losses of hadrons in compounds'', says
that if the user doesn't specify compounds, \geant treats substances as a
mixture.  Even if compounds are specified, they have to be on a short list that
doesn't include any of NOvA's major components.  So the only choice is to apply
a correction and/or systematic error after the fact.

Seltzer and Berger give a prescription for estimating $I$ and a rough error
thereon for unmeasured materials.  Hydrocarbons have been particularly
well-studied~\cite{thompson}, so this prescription is believed to be robust for
NOvA.

\begin{figure*}
\begin{center}
\includegraphics[width=0.48\columnwidth]{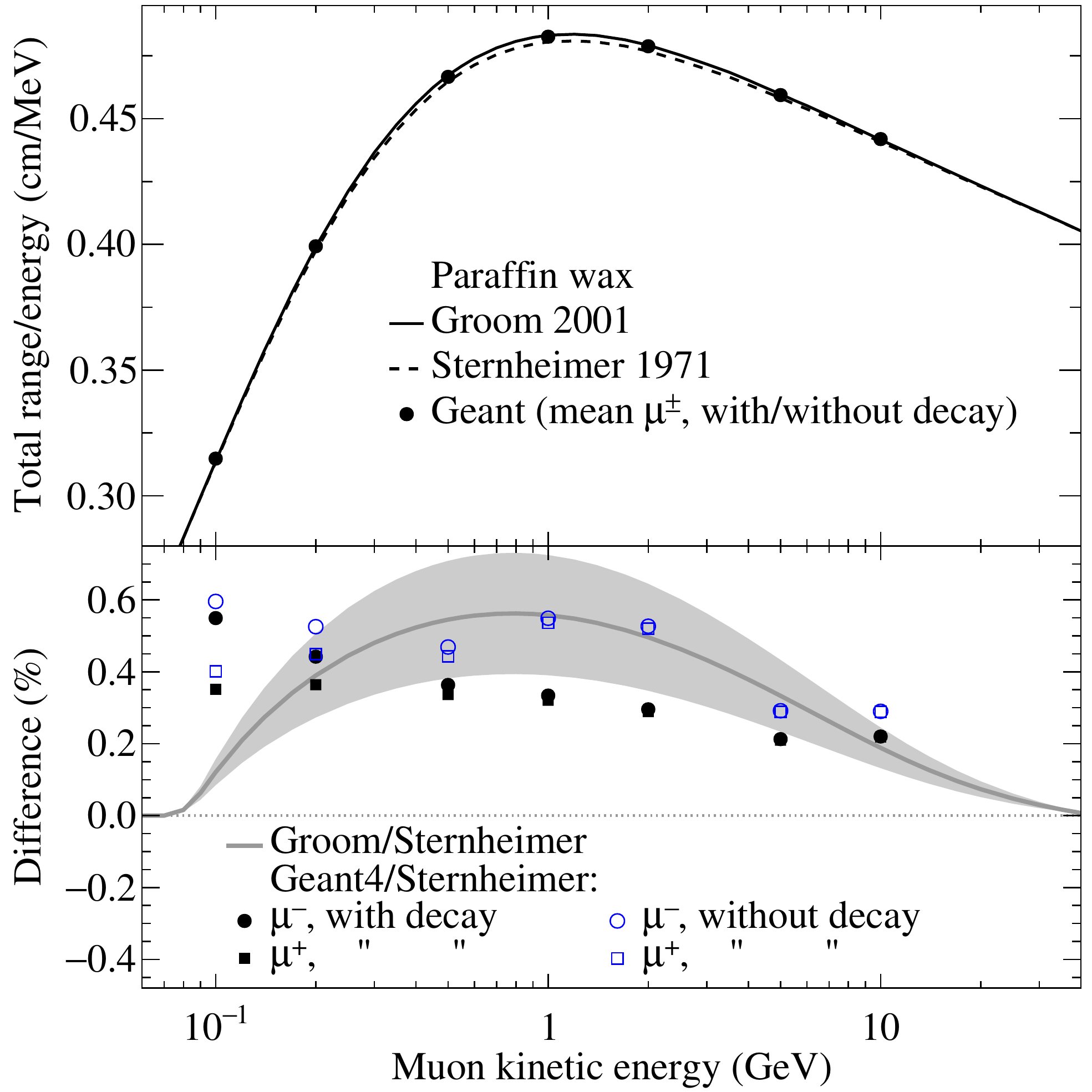}%
\hfill%
\includegraphics[width=0.48\columnwidth]{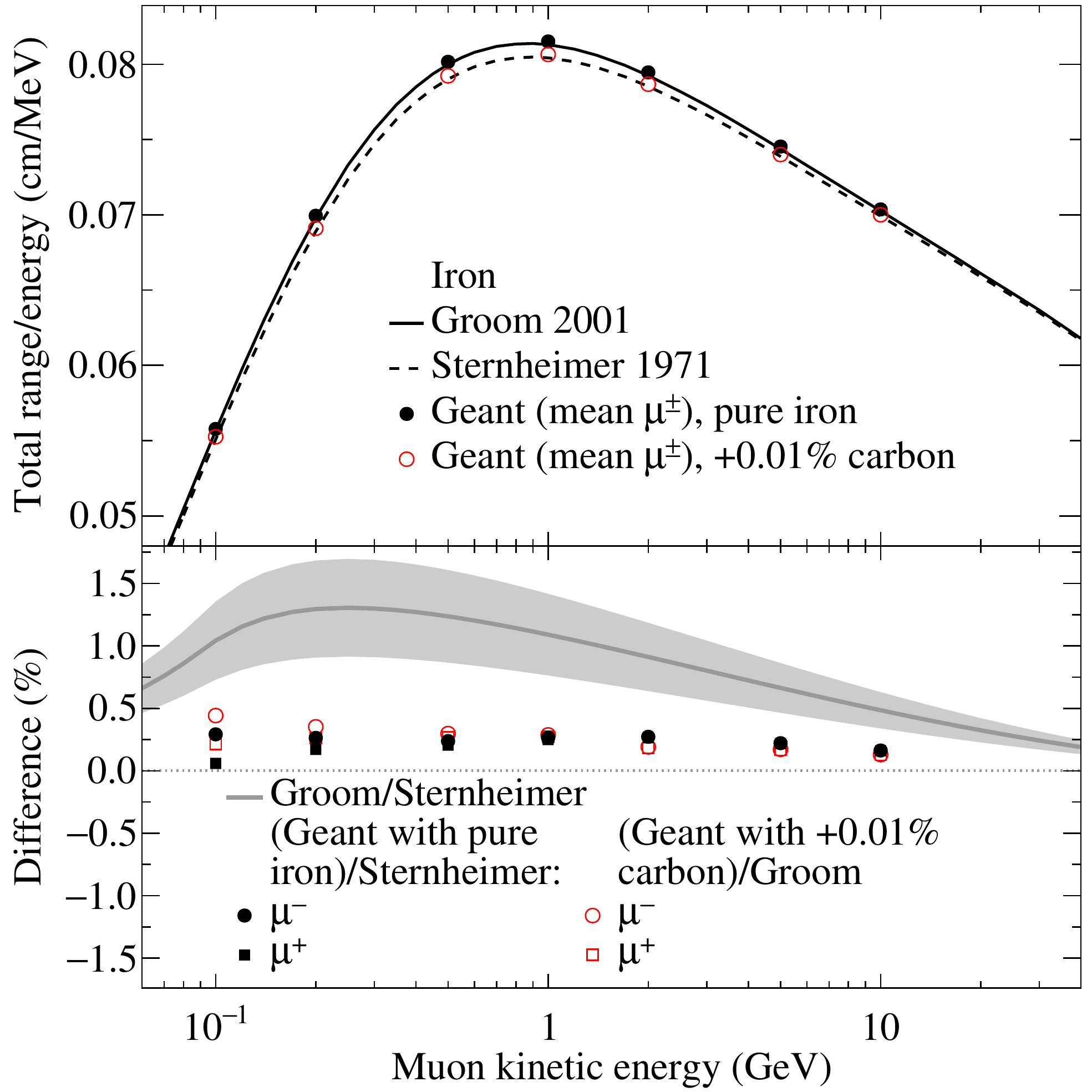}
\end{center}
\caption{\label{fig:delta}
Left (right): Comparison of the methods of
parameterizing $\delta$ in paraffin wax (iron).  The left vertical axis is the
total range (Bethe-only, ignoring shell and radiative effects) per MeV of
initial energy.  The right vertical axis is the percentage difference.  The
horizontal band shows the systematic error taken.  See
\sect{validation} for details of the \geant comparison.}
\end{figure*}

\subsection{Active planes}

Applying Seltzer and Berger's rule to mineral oil, which is itself unmeasured,
gives a shift of 0.04\% up in \dedx relative to a mixture of the same elements,
with an error of 0.15\%.

Direct measurements have been made with both PVC and titanium dioxide.  The
experimental error for PVC is 0.58\% and is consistent with Seltzer and
Berger's rule.  The rule indicates shifting \dedx down $(0.16 \pm 0.12)$\%. The rule
is therefore favored over the data.  Likewise, for titanium dioxide, the experimental
error is 1\% and is consistent with the rule, which tells us to shift down
$(0.6 \pm 0.6)$\%.

In sum, applying this rule suggests a downwards shift of $(0.06 \pm 0.14)$\%.
We take this as a systematic error of 0.2\%.

\subsection{Steel, Muon Catcher}

Molecular considerations probably do not apply to alloys~\cite[p. 204]{groom}.
In any case, since steel is nearly all iron, it cannot make very much
difference.  This leaves a \mucatchercompoundserrorrel error for the Muon Catcher
from the compounds in
active planes.

\section{Density effect, Geant4 validation}\label{sec:density}

\subsection{Density effect}

The treatment of the density effect turns out to be the biggest source of
uncertainty for all absolute and shared errors.  The PDG~\cite{pdg2016} covers
the density effect, i.e. the Bethe-Bloch term $\delta$, in Sec.~33.2.5.  While
the density effect can be calculated directly, the most popular approach, which is
implemented by \geant, uses the parameterization introduced by Sternheimer in
1952~\cite{Sternheimer1952} which approximates the effect by breaking it into
three piecewise functions of energy.  In 1971 Sternheimer 
gave a recipe for finding the values that go into this parameterization for
arbitrary substances~\cite{Sternheimer:1971zz}. This recipe restricts the
values of two of the parameters to be integers, and a third to be always zero,
for ease of computation without a computer.  This double-approximation is what
is implemented by \geant, unless
a NIST material is used.  In this case, \geant uses the values from Sternheimer
1984~\cite{Sternheimer:1983mb}.  The well-known Groom tables~\cite{groom}
use this latter set of parameters.

\geant's implementation of the density effect is 
discussed in the Physics Reference Manual in Sec.~12.1.2 on hadron-incident, and also in section 8.1.2, for
electron and positron-incident~\cite{geantmanual}.

For paraffin wax, which is a tolerable approximation of the NOvA detectors, the
1984 paper promises a deviation on the value of $\delta$ of not more than 0.052. 
The 1971 paper gives the ``average absolute maximum deviation'' for solids and liquids
as 0.178.  Roughly speaking, then, the 1984 values for paraffin wax should be about
one third as much in error as the 1971 values.

To see how different the parameterizations are, we coded up both versions for
the case of paraffin wax.  The difference is energy dependent with a maximum of
0.56\% near 800\,MeV. See \fig{delta}, which shows the results as the curves
in the top panel and their difference as the curve in the bottom panel.

For steel, we repeated the procedure, using iron as a tabulated stand-in.  The
results are also shown in \fig{delta}.  The differences are larger in iron, peaking
at 1.3\% at 250\,MeV.  While we do not use muons with a total energy of 250\,MeV (NOvA's selection fails
nearly all muons below 400\,MeV), what's
relevant is the energy of the muon when it enters the Muon Catcher, which can
easily be 250\,MeV.

If NOvA materials were among those with tabulated density effect parameters, it would be reasonable to
take an error equal to 1/3 of the difference to account for the remaining
deviation between the better approximation and the exact treatment.  However,
no substance in \nova is tabulated. 

\subsection{Validation}\label{sec:validation}

It is worth checking whether \geant's implementation of the Bethe-Bloch formula
results in the correct range, with the best estimate of that correct range
given by Groom 2001~\cite{groom}.  There is some expected deviation from the use of the Sternheimer
1971~\cite{Sternheimer:1971zz} parameterization of the density effect, as
estimated in \sect{density}.  It is also expected that the \geant range will be
very slightly too long because it doesn't know our elements are in compounds
(see \sect{compounds}), but this is only 0.04\% for the active planes and less
in the Muon Catcher.  Groom does not mention muon decay in flight, so it is
unclear whether it counts as a source of energy loss or not; this matters at
the O(0.1\%) level.  Finally, Groom neglects the difference of ``a few parts
per thousand'' in the range between \mum and \mup, but \geant does not.  For
\mum (\mup), we can expect the \geant range to be slightly longer (shorter)
than the Groom range.

We ran many $\mu^-$ and $\mu^+$ through a detector model made of solid paraffin
wax another made of solid iron, and a third made of 99.99\% iron with 0.01\%
carbon.  The motivation of the last was to test \geant's switch from a
tabulated to untabulated material.  Our implementation of paraffin wax is
$\mathrm{C_{25}H_{51}}$, motivated by Groom's definition
$\mathrm{CH_3(CH_2)_{n\approx23}CH_3}$.  This has one fewer hydrogen than
Groom's, assuming the intent was for it to be strictly saturated.  The
definition of paraffin wax in Sternheimer 1984 includes the $Z/A$, and all the
parameter values match Groom's.  This $Z/A$ implies $\mathrm{C_{25}H_{51.7}}$.
The difference between 51 and 52 makes a difference of 0.02\% in the ratio of
ranges for the two density effect parameterizations, with the
singly-unsaturated hydrocarbon being slightly more different.  It makes much
less difference whether $\approx 23$ is interpreted as 23 or some other nearby
number.  The reference density of 0.93\,g/cc was used.  Iron was simulated in
the same way, and at 7.874\,g/cc to match the Groom tables.  

Care was taken to avoid ambiguities in the definition of track length.
\geant gives access to the physical path length,
via \texttt{G4Step::\allowbreak GetStepLength()}.  This length includes the deviations
from multiple Coulomb scattering.
For comparison with Groom, we believe
it is correct to compare with this length.  Groom says that ``multiple
scattering is neglected,'' which we take to mean that ranges are the full
physical length, not the (slightly shorter) depth that a particle penetrates
into a block of material.  This isn't completely clear, and it's also doubtful
whether the various old experiments that provide all the data for parameters
such as the mean excitation energy made this distinction.  For that matter, as discussed in
\rf{GOTTSCHALK1993467}, there's been confusion over the years as to what theory
should be used for multiple scattering.

\begin{figure}
\begin{center}
\includegraphics[width=\columnwidth]{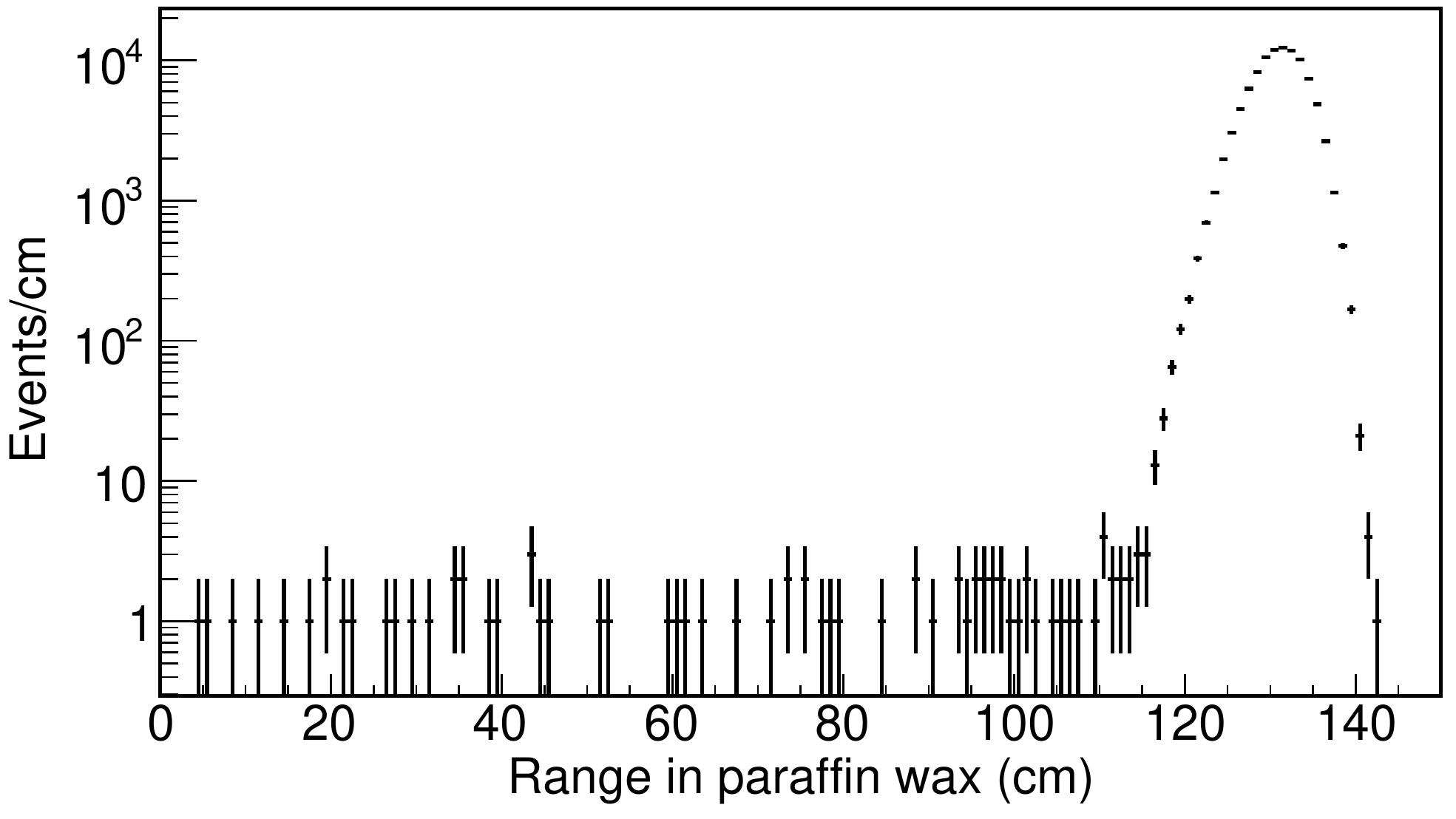}
\end{center}
\caption{\label{fig:range} 
\geant range example: 300\,MeV $\mu^-$ in paraffin wax, which is a stand-in
(tabulated in Groom 2001) for the plastic and oil of the NOvA detectors.
This illustrates the point that the PDG (and others) make about how the
most likely range is very different from the mean range.  ``Range'' here is the
true path length.
}
\end{figure}

Following the cautions in many sources~\cite{pdg2016,groom,Bichsel:2006cs}
about how many events must be simulated to obtain a reliable mean \dedx, 
10$^5$ events were simulated in each case.  An example distribution is
shown in \fig{range}.

\fig{delta} shows the results of the validation study.  Because of the
ambiguities noted above, the bottom panels show the
difference between the expected range, using the Sternheimer 1971
approximation, and \geant for both \mum and \mum and both with and without muon
decay.  \geant gives consistently longer ranges than expected, with only a
slight energy dependence.  Since the ranges are 0.2--0.6\% longer than expected
from the Sternheimer 1971 approximation, they are quite close to the Groom
ranges.  But this must be an accident since (a) \geant does not know that this
is paraffin wax so cannot look up the parameters for paraffin (b) the shape
shown in the bottom panel does not track the difference between Sternheimer and
Groom.

For iron with a tiny amount of carbon added to prevent \geant from using the Sternheimer 1984 parameters
for this tabulated substance (also used by Groom), the range also tracks Sternheimer 1971 as expected, but is also a little high, with
the mean of \mum and \mup being 0.1--0.3\% above what's expected.  The difference between
allowing or disallowing muon decay as energy loss is not plotted in \fig{delta} because
it makes so little difference for iron, given its high density.
For pure iron, the expected shift to tabulated parameters is seen.  Now
the range tracks Groom instead, while \emph{also} being 0.1--0.3\% high.

\subsection{Uncertainty taken}

The situation for the FD and main ND is as follows:

\ul

  \item \geant uses an old approximation that causes the range to be 
        0.6\% too short around 1\,GeV for a reference hydrocarbon that is similar
        to the primary detector materials.

  \item The error on the new approximation can be estimated to be 0.2\%
        at 1\,GeV, but since neither NOvA scintillator nor
        plastic is a tabulated material, it isn't possible to use it.

  \item Some unknown feature of \geant causes the range to be more-or-less uniformly
        $\sim$0.4\% longer than expected (i.e. the Groom range when
        using the 1984 parameters or the same modified by the difference in
        density effect parameterizations when using the 1971 parameters).  This
        could be a bug, or it could be that \geant takes into account real
        effects that Groom neglects.

\lu

So in the range 400\,MeV to 4\,GeV which is relevant for most NOvA analyses,
\geant (apparently) accidentally agrees quite closely with Groom, but if 
Groom and \geant are assumed to use equally good code, then we have to estimate a 0.4\% error
from two attempts to model the same process coming up with different answers. 

There is also the 0.2\% error on the true size of the density effect from the better
approximation. 
That error is almost certainty anti-correlated between energies within the
relevant range for NOvA, since the three-part piecewise Sternheimer
approximation is meant to keep the result close to the exact form and therefore
probably crosses it at least twice.  Without undertaking the exercise of
calculating the exact form, it's not clear what there is to be done about this.
Another complication is that the exact form is still subject to experimental
error, namely it is a function of the mean excitation energy
$I$~\cite{Sternheimer1952}.

The situation for the Muon Catcher steel is similar except there is no
accidental agreement.  We have decided
to assign a density effect+validation error equal to the quadrature sum of:

\ul

  \item The greater of:

    \ul

      \item The difference between the two sets of density effect parameters (given
      that \geant uses the worse ones), and

      \item The energy-dependent difference between \geant and the
            Groom calculation, averaged over \mup and \mum, and over
            including muon decay and omitting it.

    \lu

    The former is greater at all energies used by the $\nu_\mu$ disappearance analysis.

    \item Twice the error on the 1984 approximation, taken as 0.3 times the difference
          between the two sets of density effect parameters (0.6 times after doubling).
          It is doubled to make some attempt to account for the complications discussed
          above.

\lu

The result is then scaled up by 1.1 for the main detector, to attempt to cover
the chemical differences between paraffin wax and the detector materials, and by 1.01 for
the steel, since iron and steel are very similar. 


The main detector does not have the same density as the reference paraffin wax,
but at a mean of 0.98\,g/cc, it is quite close.  The density effect is somewhat
stronger for the PVC and somewhat weaker for the scintillator.  Numerically,
the cancellation is nearly perfect, so the magnitude of the effect in paraffin
wax is the same as the mean magnitude in the detector.  Likewise, no scaling of
the effect is needed for the steel since the reference density is within half a
percent of our steel planes.

This results in a maximum
error, from the density effect and \geant validation, of 0.8\% for the main
detector and 1.2\% for the Muon Catcher.  The density effect/validation
error from the active detector and the steel are taken to be uncorrelated.

\subsection{Inhomogeneities}

The above discussion assumes homogeneous mixtures with
constant density.  It is not clear how to treat the density effect if the
material is microscopically inhomogeneous~\cite{icru37}.  This is true for
NOvA plastic, which contains 15\% powdered titanium dioxide (anatase crystalline
form) by weight~\cite{Talaga:2016rlq} with density
3.78\,g/cc~\cite{wikipediaanatase} mixed with the PVC itself which has a
density 1.1\,g/cc.

Using the same method as \rf{icru37}, the severity of this effect is determined
by evaluating the density effect for the case of a uniform material and one
split entirely into two density regions.  The \dedx in the latter case is
lower. Modeling the PVC as paraffin wax again, the \dedx difference at 1\,GeV, which
is approximately where the effect is largest, is 0.33\%.  Since the real effect
must be between zero and this, presumably depending on how the maximum impact
parameter for excitations compares to the size of the titanium dioxide grains,
the midpoint is taken as the best estimate and the error as a uniform
distribution with RMS $0.33\%/\sqrt{12} = 0.10\%$. The error is on the PVC
only, making a shift in \dedx of $(-0.05 \pm 0.03)\%$ overall.

\coulombsection{1}

\begin{figure}
\begin{center}
\includegraphics[width=0.95\columnwidth]{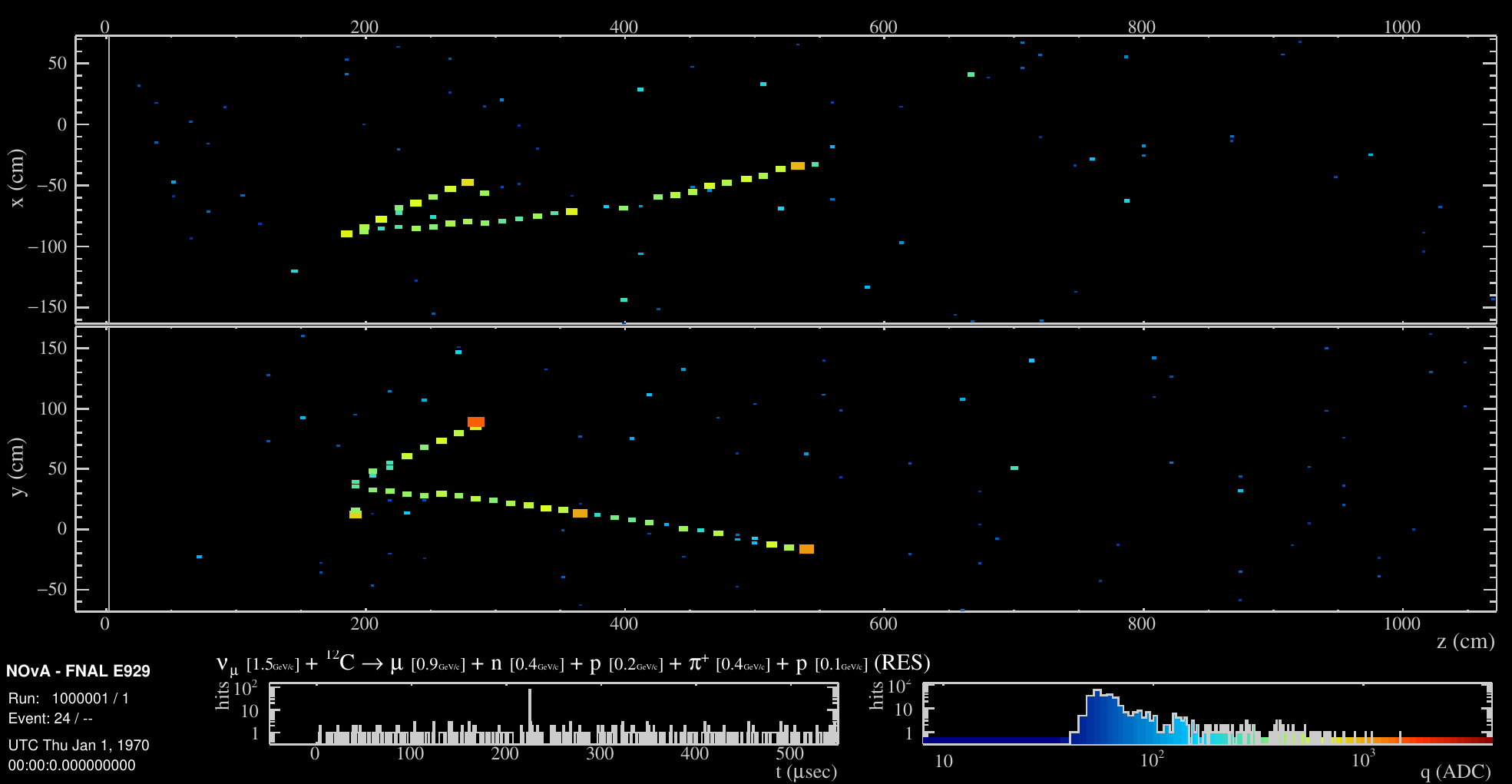}

\includegraphics[width=0.95\columnwidth]{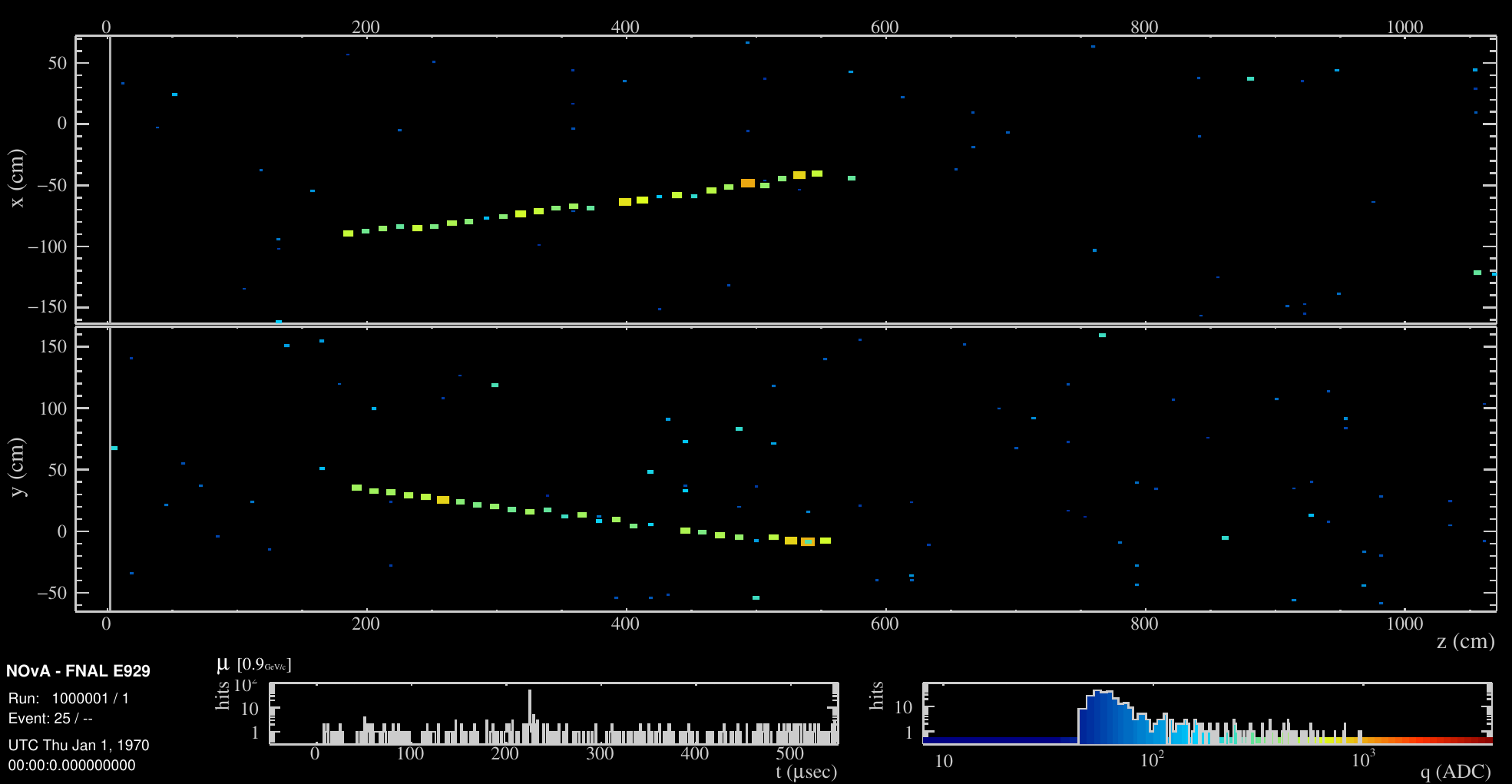}
\end{center}
\caption{\label{fig:ho24}
An event and just its muon resimulated, with 
multiple scattering being the only significant difference. The reconstructed muon track is
3\,cm longer in the complete event. }
\end{figure}

\section{Hadronic modeling and response}\label{sec:hadronicoverlap}

Inaccuracies in hadronic modeling could cause the data and MC to have different
amounts of overlap between the muon and hadronic system which could cause
the reconstructed track length to differ in an unmodeled way.  This could further
be somewhat different between the ND and FD because of the different
acceptances.  Unlike the previously discussed sources of error, this
error is dependent on details of the neutrino interactions in the NuMI
beam.  The standard neutrino-mode NOvA beam is assumed throughout.

\begin{figure}
\includegraphics[width=\columnwidth]{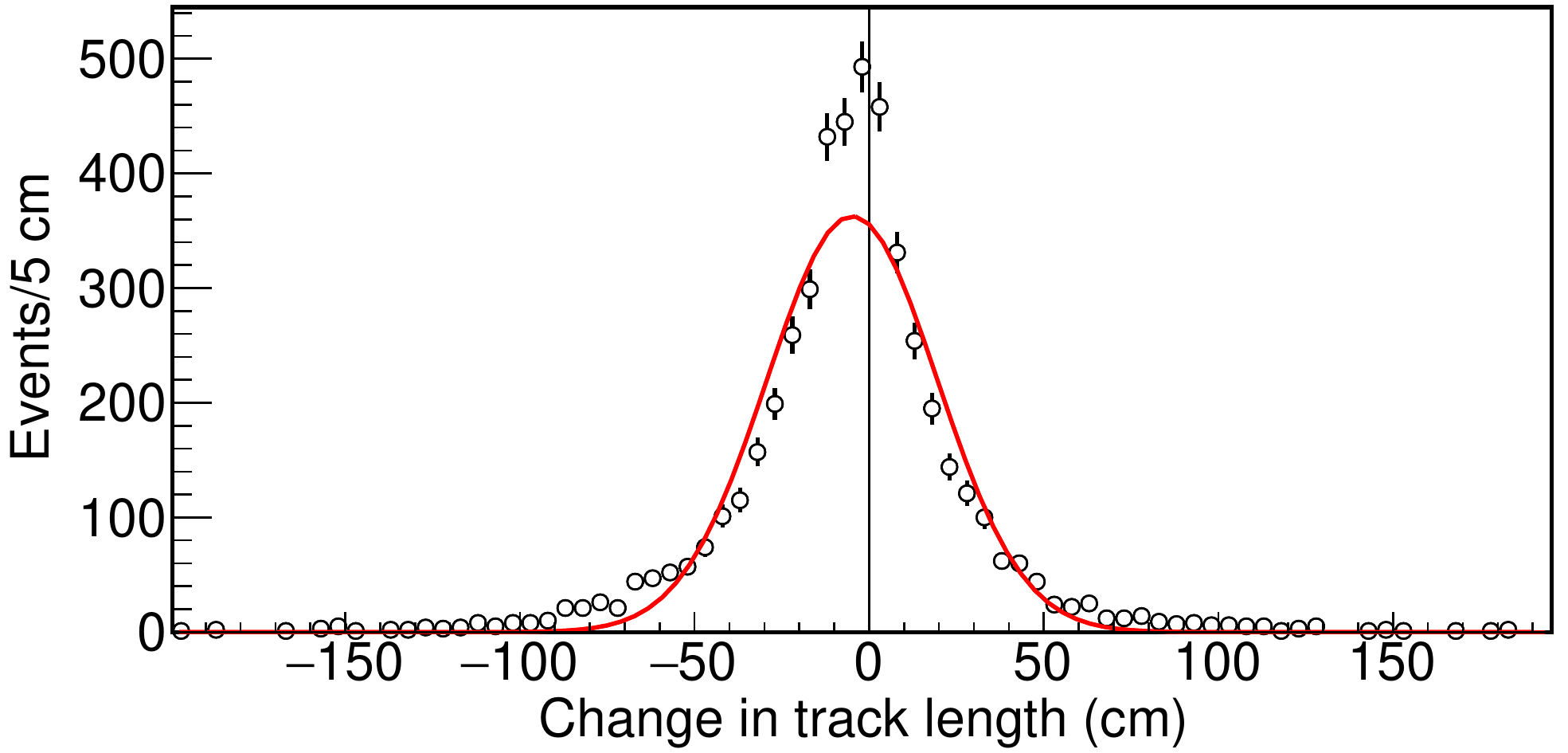}
\caption{\label{fig:hadronicoverlap} \protect\input{mufig/hadronicoverlap.txt}}
\end{figure}

To put a bound on the size of this effect, a study was performed
to check how much reconstructed
track length changes for neutrino events if all primary particles other than
the muon are removed from the event after \genie~\cite{genie} and before \geant.
The muons are resimulated in this study, 
and so some change is expected from real differences in the muon track, but
this change is zero on average.  A sample of 
$5000$ ND events with well-contained muons was used. The
reconstructed track length differences are shown in \fig{hadronicoverlap}.

An example of the procedure is shown in \fig{ho24}.  The original event is
\[\nu_\mu\ ^{12}\mathrm C\ \overset{\mathrm{RES}}{\rightarrow}\ \mu^-\
^{9}\mathrm{Be}\ \mathrm n\ 2\mathrm p\ \pi^+\] at 1.5\,GeV.  Note how in the
muon-only event the muon does not follow exactly the same trajectory.  In this
case, the original reconstructed muon track is 363.7\,cm long and the bare muon
track is 360.9\,cm long.  The bare track \emph{looks} one plane longer in the
event display, but the last hit in each view is from the Michel electron and is
not in time with the track, whereas only the last $x$ hit is a Michel in the
full event.  The reconstructed start and end points in $z$ are the same for
each version of the event, but the bare muon travels slightly less distance in
$x$ and $y$ than the muon in the full event, i.e. in this case, the difference
in length is a function of different multiple scattering.

\begin{figure}
\begin{center}
\includegraphics[width=0.95\columnwidth]{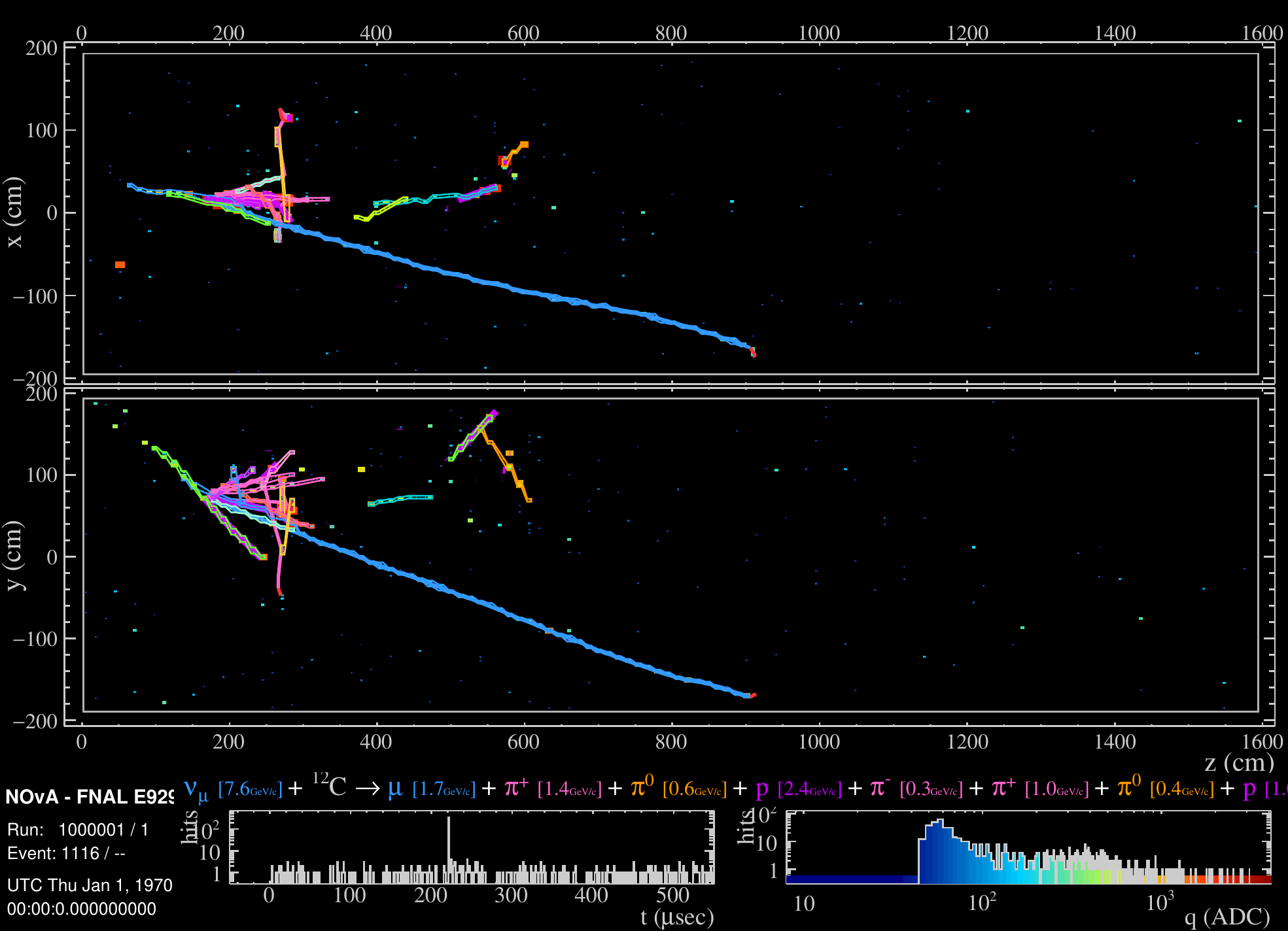}

\includegraphics[width=0.95\columnwidth]{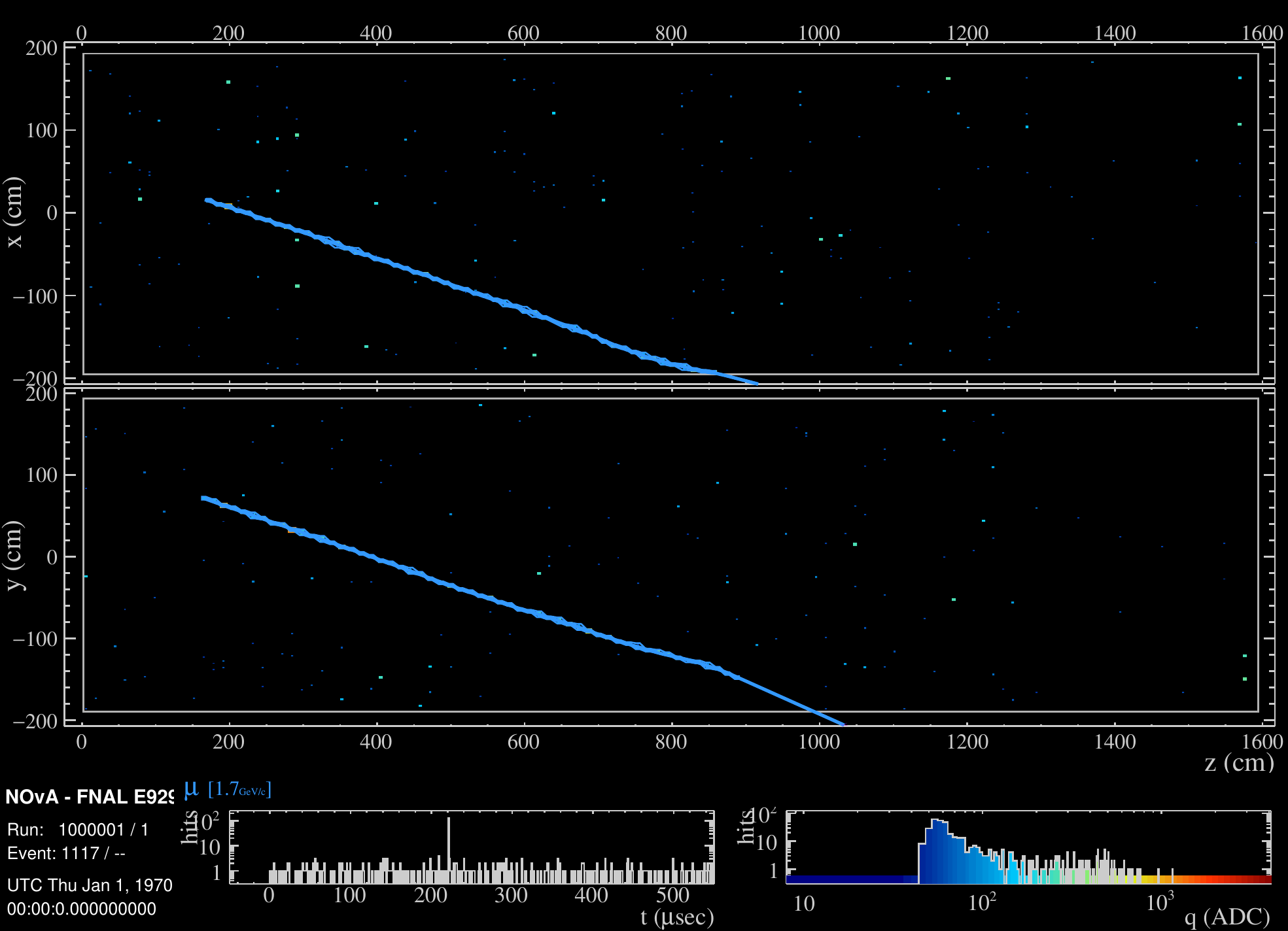}
\end{center}
\caption{\label{fig:ho1116}  An event in which the tracker attaches some hits
from a backwards-going particle to the forward-going muon track, and 
the muon resimulated alone.}
\end{figure}

A more complex example is shown in \fig{ho1116}.  \[\nu_\mu\ ^{12}\mathrm C\ 
\overset{\mathrm{DIS}}{\rightarrow}\ \mu^-\ ^{10}\mathrm{B}\ 2\mathrm p\ \mathrm 2\pi^+\
2\pi^0\ \pi^-\] at 7.6\,GeV.  One of the \piz is backwards-going
and the reconstruction has incorrectly connected the muon track to its hits.

\newcommand{\overallhadronicshift}{5.6}
\newcommand{\overallhadronicshifterror}{0.5}
\newcommand{\overallhadronicshiftpercent}{1.94}
\newcommand{\overallhadronicshifterrorpercent}{0.18}

The average track is $(\overallhadronicshift \pm
\overallhadronicshifterror)$\,cm --- or $(\overallhadronicshiftpercent \pm
\overallhadronicshifterrorpercent)$\% --- shorter by itself than it is with the
whole event.  About 70\% of this shift is due to small changes in track length,
visible as the shift in the peak from zero. Evidently on average there is some
soft backwards-going shower activity that gets incorrectly attached to the
track, typically one plane's worth.  Or alternatively, the presence of several
charged particles in the first cell cause that cell to be over threshold
whereas with just the muon --- traversing part of a cell --- it would not be.
The other 30\% of the shift is due to substantial changes that form the
asymmetric tail on the low end in \fig{hadronicoverlap}, presumably mostly
originating from backwards-going particles that get attached to the muon track.

In the following discussion, the fraction of this \overallhadronicshiftpercent\% 
that needs to be taken as a systematic error is evaluated.

\subsection{Overall cross sections}

\begin{table}
\begin{center}
\begin{tabular}{l | c c | c}
\hline
\hline
Interaction & $\Delta l$ (cm) ND & \% ND & \% FD\\
\hline
Coherent       & $1\phantom{.0} \pm 4\phantom{.0}$ & $-1.5 \pm 1.5$           & $1.0\phantom{0} \pm 0.3\phantom{0}$ \\
Quasielastic   & $1.5 \pm 1.2$                     & $\phantom{-}0.9 \pm 0.3$ & $0.94 \pm 0.07$ \\
MEC            & $3.5 \pm 1.4$                     & $\phantom{-}1.1 \pm 0.3$ & $1.17 \pm 0.09$ \\
Resonant       & $7.1 \pm 0.7$                     & $\phantom{-}2.9 \pm 0.2$ & $2.29 \pm 0.06$ \\
Deep inelastic & $7.9 \pm 0.9$                     & $\phantom{-}1.7 \pm 0.5$ & $2.04 \pm 0.09$ \\
\hline

All & $ \overallhadronicshift        \pm \overallhadronicshifterror       $
    & $ \overallhadronicshiftpercent \pm \overallhadronicshifterrorpercent$ 
    & $1.78 \pm 0.04$ \\
\hline
\hline
\end{tabular}
\end{center}
\caption{\label{tab:hadronicoverlap} Increase in reconstructed Monte Carlo muon track length
caused by the non-muon part of the event. As expected, event types with more hadronic activity
show more change.}
\end{table}

This difference is much larger than would be caused by any mismodeling of
neutrino interactions. There \emph{is} hadronic activity; the only question is
the details.  To get a better handle, the sample is broken into coherent,
quasielastic, resonant, MEC, and deep inelastic events.  It is also repeated
for the FD. See \tab{hadronicoverlap}.  All of the results are consistent
between ND and FD except for resonant events, which are reconstructed with less
of the hadronic activity attached to the muon track in the FD.  Since the other
interaction types are consistent, this is believed to be a flux difference
rather than a detector effect.

The smallest difference between track length with and without hadronic activity
is seen for quasielastic events, which stands to reason. Some idea of a
reasonable upper bound on the systematic uncertainty can be obtained by
considering the change when each interaction type is turned off entirely.
Without MEC, the shift is 0.07\%.  Without resonant events, there is a 0.29\%
shift at the FD and 0.7\% at the ND. Or without quasielastic, 0.25\%.  A
conservative estimate for the relative uncertainty in the cross sections
between interaction types is perhaps 30\%~\cite{McGivern:2016bwh}, making a
reasonable systematic error from cross sections 0.23\% (ND) and 0.11\% (FD).
The shape of the cross sections matters as well as the overall normalization,
but given such a large normalization error, this is neglected.

Even these small errors could be eliminated if the muon energy
estimator --- the reconstruction code that translates observed muon track length
to muon energy --- were retuned at each step of the evaluation of cross section errors in the
overall oscillation analysis.  For simplicity, this is not done. The error
on muon energy from cross sections
arises from using the only nominal Monte Carlo to tune the energy estimator.

\subsection{Topology}

Unmodeled or mismodeled event topology must also be considered. To constrain
mismodeling in which \genie predicts the wrong distribution of backwards-going
charged particles that leave significant tracks, a low-statistics
scanning study was performed.  Looking at 50 contained Near Detector data and MC events, it was 
noted how many times there seemed to be a track of at least two planes coming
backwards out of the neutrino vertex.  The assumption is that the total number
of these is proportional to the number that are sufficiently in line with the
muon that the reconstruction attaches them together.

In each
case, there were nine such events.  This constrains the data/MC ratio for such
events to $1.0 \pm 0.5$.  As given above, events where a track like this gets
attached to the muon track contribute about 30\% of the mean shift.  So this
gives an uncertainty of 0.5 of 30\% of \overallhadronicshiftpercent\%, or
0.26\%.

In each case discussed so far, for purposes of reconstructing neutrino energy,
any hit which is transfered from the shower to the track still contributes
to the total energy.  The only difference is whether it contributes towards an
estimate of muon energy by track length or hadronic energy by calorimetry.
Roughly, the average hit contributes 13\,MeV if it is part of a muon track and
17\,MeV as part of a shower, so the shift only changes the total event energy
by 24\% of the amount the track energy shifts.  This brings the effective
systematic from cross-section uncertainties and topology, taken as
uncorrelated, to \modeledhadronicoverlapabserror.

This study was done with NOvA's Kalman tracker, which knows nothing of the
neutrino interactions and will follow a track
through the neutrino interaction vertex.  A newer tracker called
Break\-Point\-Fitter is aware of the vertex and will not connect particles
going backward with those going forward, provided the vertex is correctly
identified.

\subsection{Remnant nucleus}\label{sec:boiloff}

An unmodeled effect that does \emph{not} simply represent a shift from track to
shower energy is that \genie does not model boil-off nucleons or de-excitation
photons from the remnant nucleus\footnote{Except if the struck nucleus is
oxygen it does emit photons, because Super-Kamiokande needed that~\cite{genie}.} So there is a
population of unmodeled $\sim 1$--$10$\,MeV isotropically emitted particles
that can add a hit or two near the vertex and shift the muon energy.

Let us suppose that the remnant nucleus on average emits one proton of 5\,MeV,
somewhat above carbon's Coulomb barrier, and one photon of 5\,MeV, typical of
de-excitation.  The proton's average range is 0.3\,mm~\cite{pstar} and so only
matters in so far as it might boost the cell hit over threshold.  After Birk's 
suppression,
the effective energy is about 0.5\,MeV. Given that 62\% of the neutrino
interactions are in scintillator, and that a muon that crosses the whole cell
deposits 12\,MeV, it may push the hit over threshold 3\% of the time.  This
effectively lengthens the average track by 0.2\,cm.

The photon, in contrast, has a $\sim 20$\,cm interaction length and will
deposit $\sim 3$\,MeV on average the first time it interacts.  The probability
of it going backwards and reasonably in line with the track in 2D is about 25\%,
and the probability of interacting in that plane is about 25\%, so it may
extend the track about 6\% of the time.  Putting this together with the
estimate of the effect from protons, a reasonable estimate for the lack of
de-excitation in \genie is that tracks in data are $\sim 0.6$\,cm longer than in
MC.

Putting these together, and noting that if a harder particle scatters
backwards, which happens about 20\% of the time (see above), these nuclear
effects don't matter, this represents an unmodeled increase in track length of
about 0.5\,cm, or 0.17\%.  Added to the errors above, this gives \hadronicoverlapabserror for the 
absolute error in either Near or Far Detector due to how tracking 
interacts with hadronic modeling.

\subsection{Near/Far}

The beam spectrum is different between the Near and Far Detectors, 
the acceptance of the Near Detector is much more limited, and the hit threshold
at the Far Detector is higher. However, the cross section uncertainties mostly
cancel and the other effects are mostly covered by the MC, so one must be left
with a small fraction of the absolute error, which is already small.
To be concrete, the residual error is assumed to be a quarter of the absolute, 
making it \hadronicoverlaprelerror.

\section{Near Detector neutron pile-up}\label{sec:neutronpileup}

\muoncatcherneutrontext{1}

\section{Negligible effects}\label{sec:dontmatter}

Uncertainties from noise modeling,
muon decay, and multiple scattering are each found to have an effect of order
0.005\%, while block and plane alignment has an effect of about 0.01\%.

\subsection{Noise modeling}

Differences in detector noise between data and MC could result in additional
hits being added to tracks in an unmodeled way.  Since the Near Detector is 
much quieter than the Far Detector, this would be mostly a 
FD effect.

\begin{figure}
\includegraphics[width=0.96\columnwidth]{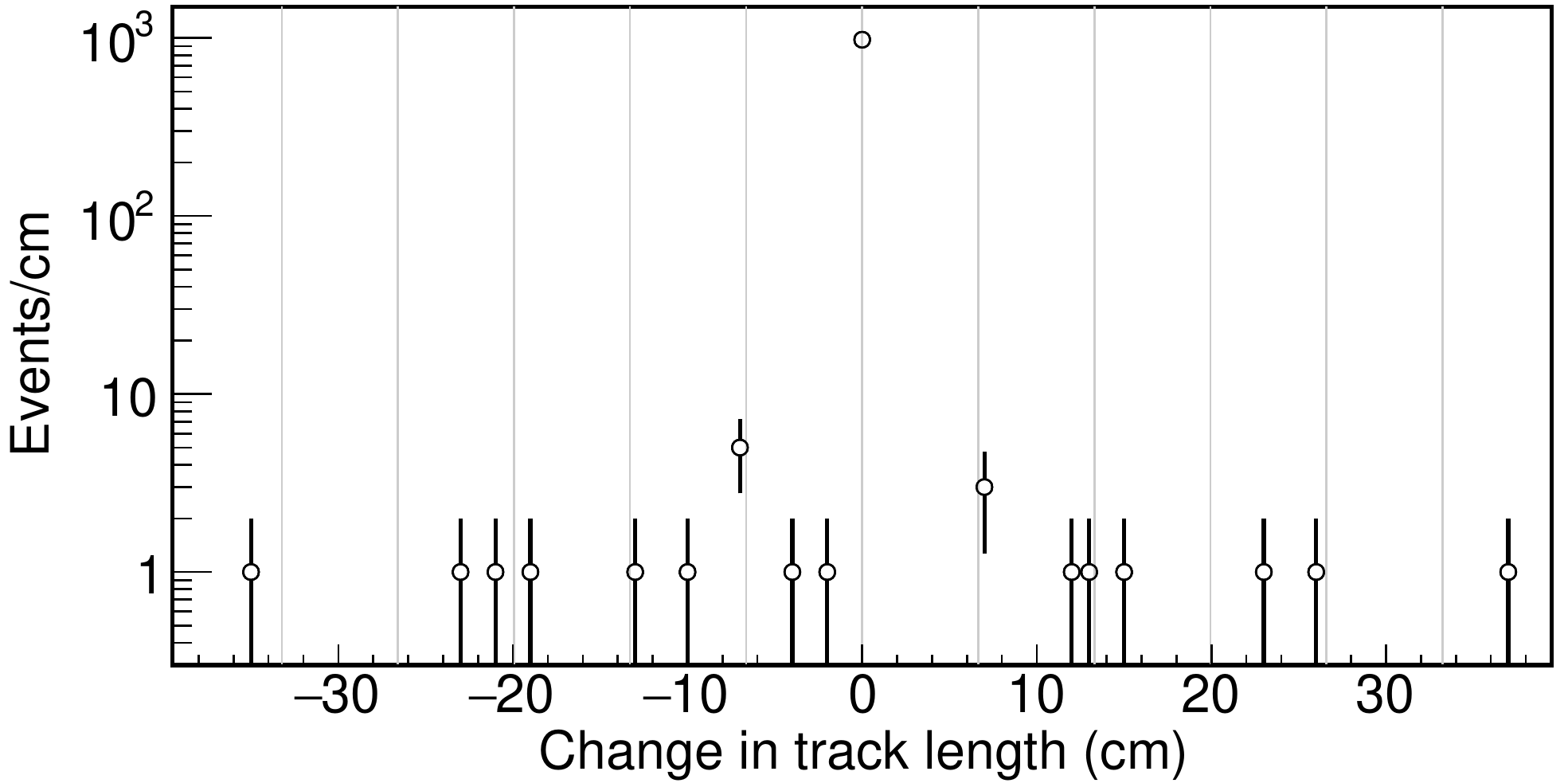}
\includegraphics[width=0.96\columnwidth]{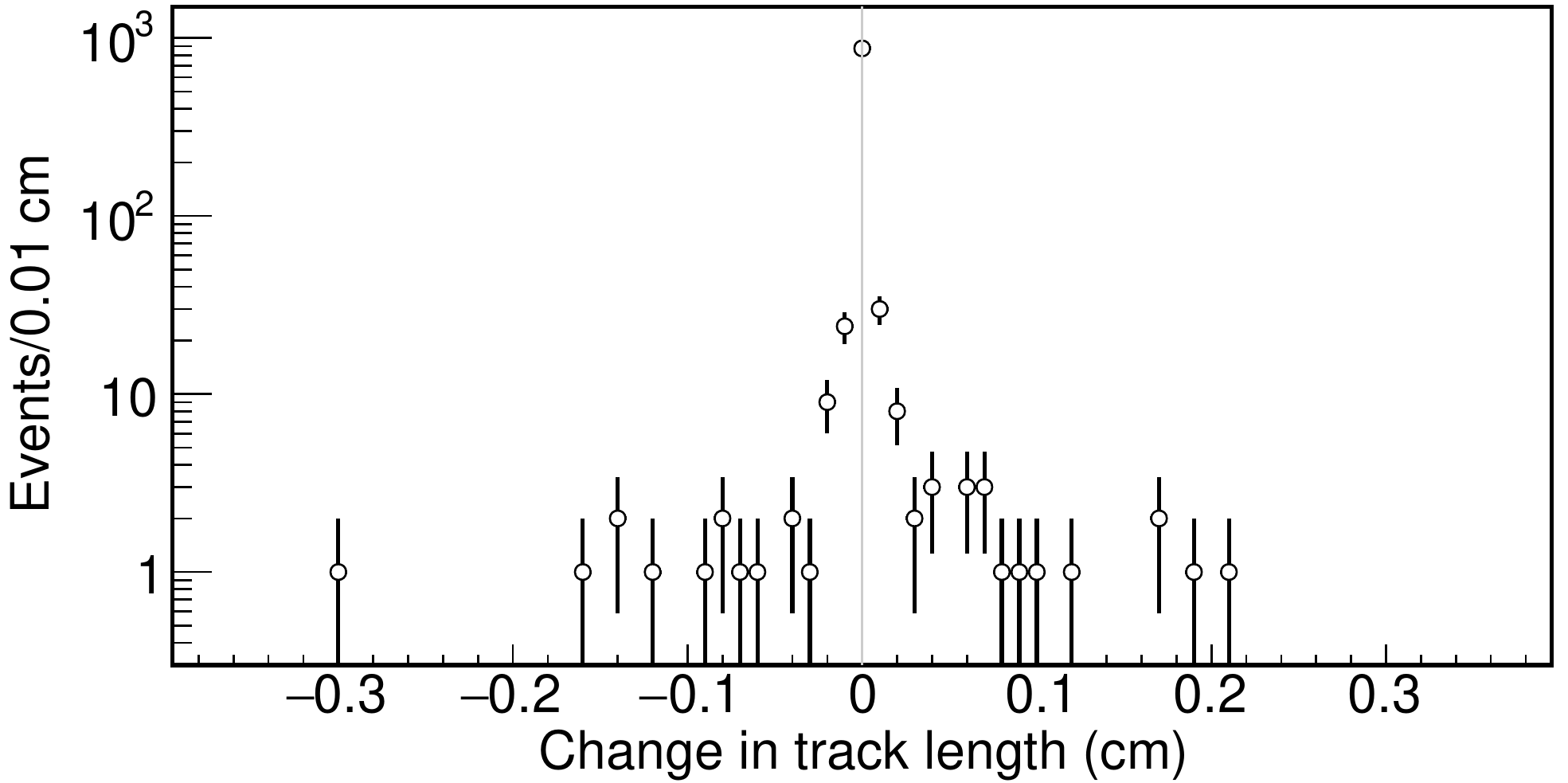}
\caption{\label{fig:overlay}
Changes in track length with cosmic
overlay. Top: from $-40$\,cm to $+40$\,cm.  The gray lines show plane depths; the most common
significant change is for one plane to be added or subtracted. Bottom: zoomed in on $-0.4$\,cm to +0.4\,cm,
showing that the vast majority are not changed at all.}
\end{figure}

In the course of NOvA analysis, several different approaches to handling noise
have been used.  As the effect will be found to be very small, let us make an
estimate bounding the magnitude of a potential problem with noise modeling that
is agnostic to the approach.  One thousand 2\,GeV muons were
simulated in the FD and overlaid with cosmic data.  The reconstructed track length of simulated muons
was compared with and without the cosmic overlay.  Nearly half of the tracks
changed length, but most changes were less than 0.3\,cm, caused by additional
hits along the track slightly changing the trajectory, with 97.6\% changing by
less than this amount.  The remaining 3.4\% changed by up to 40\,cm (see
\fig{overlay}).  

\begin{figure}
\begin{center}
\includegraphics[width=0.95\columnwidth]{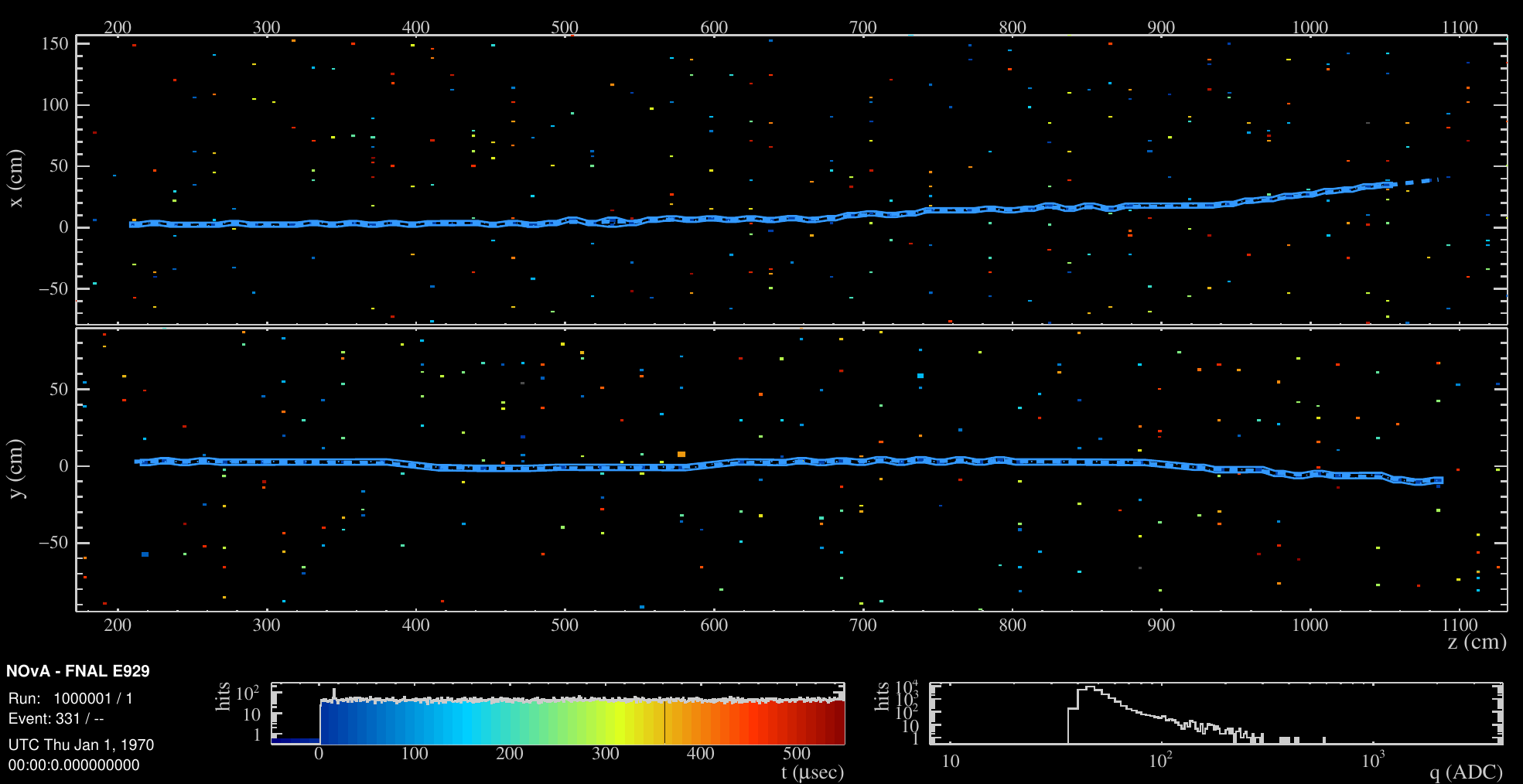}

\includegraphics[width=0.95\columnwidth]{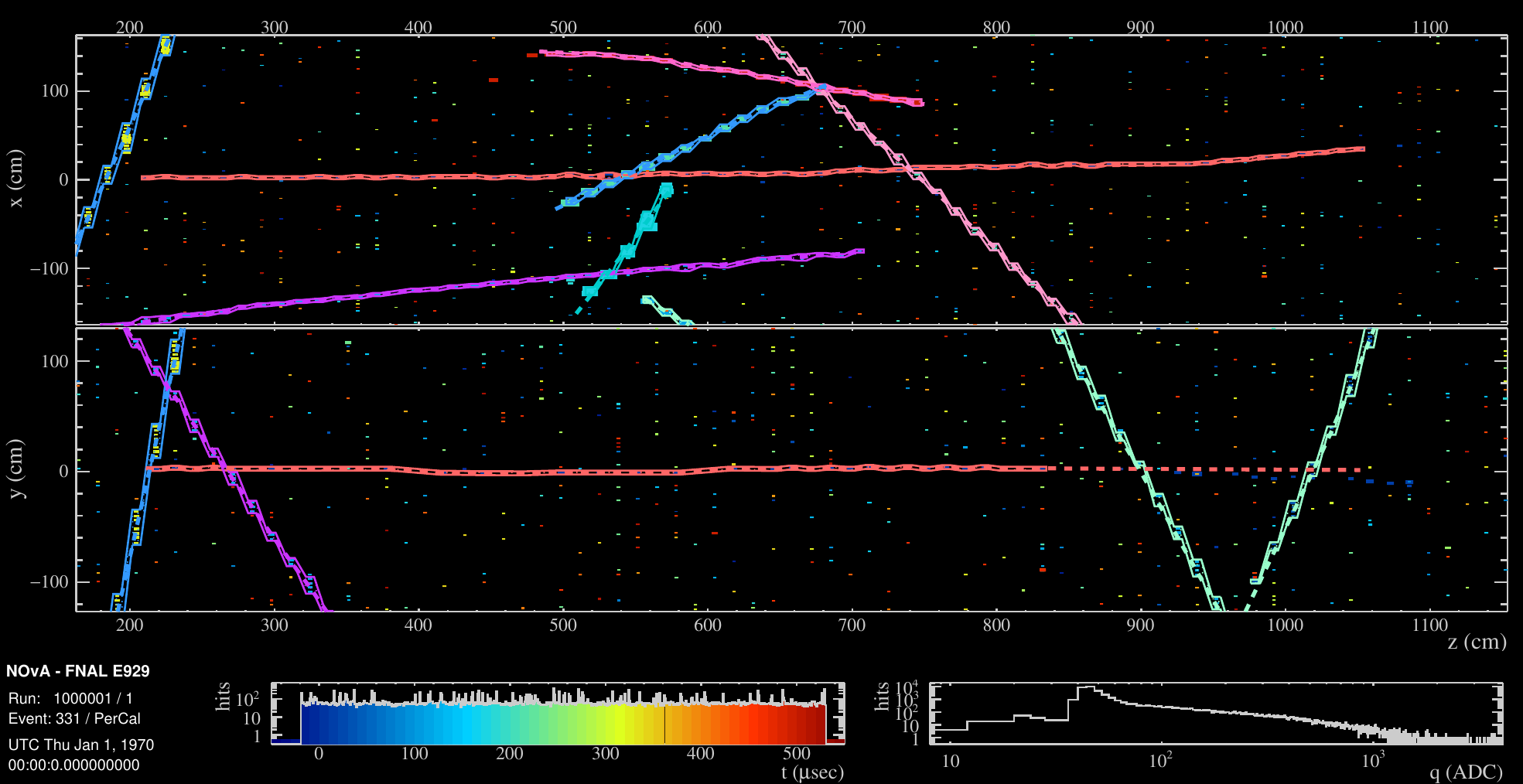}
\end{center}
\caption{\label{fig:mcmuon331} Example of a track shortened by cosmic overlay. }
\end{figure}

\begin{figure}
\begin{center}
\includegraphics[width=0.95\columnwidth]{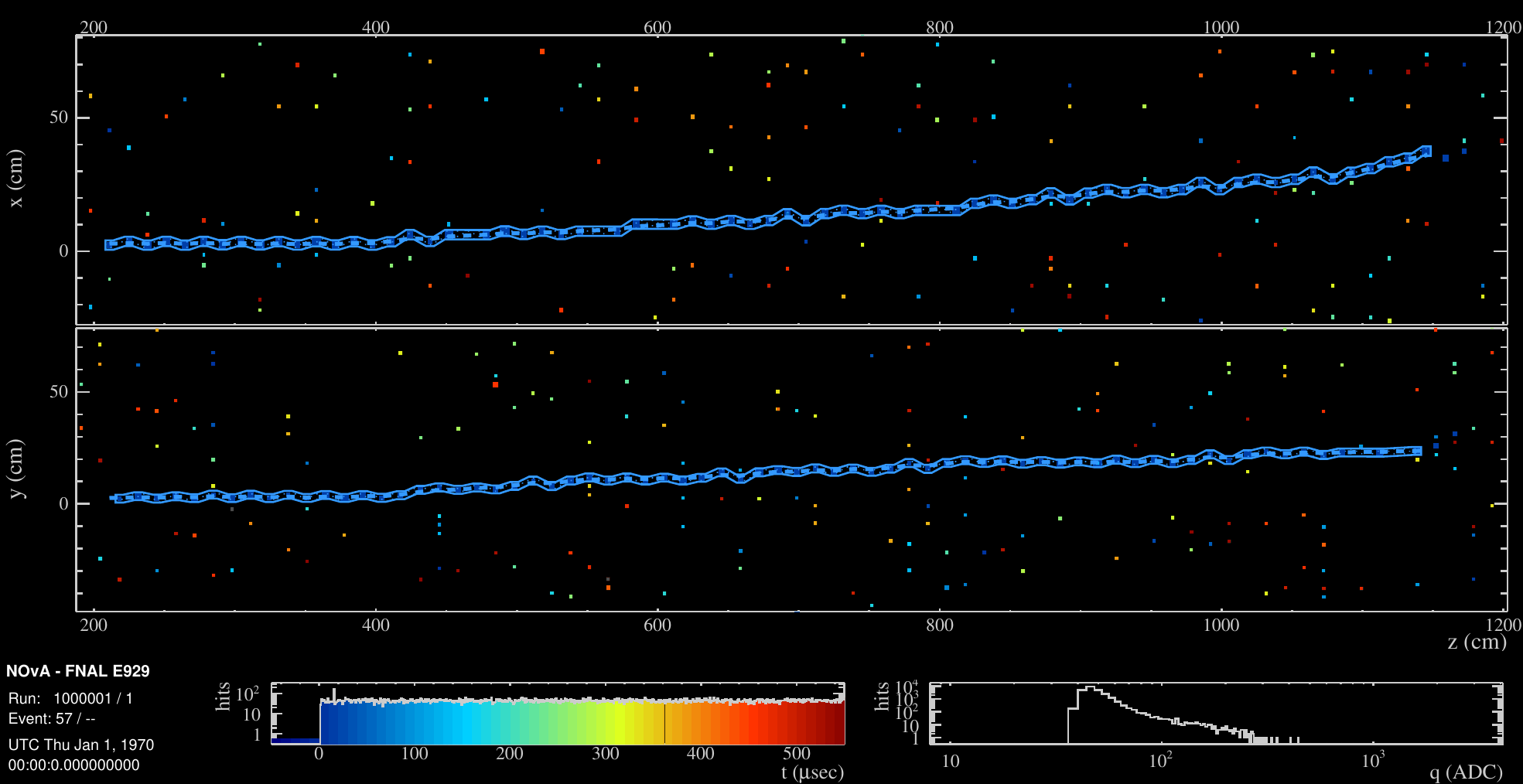}

\includegraphics[width=0.95\columnwidth]{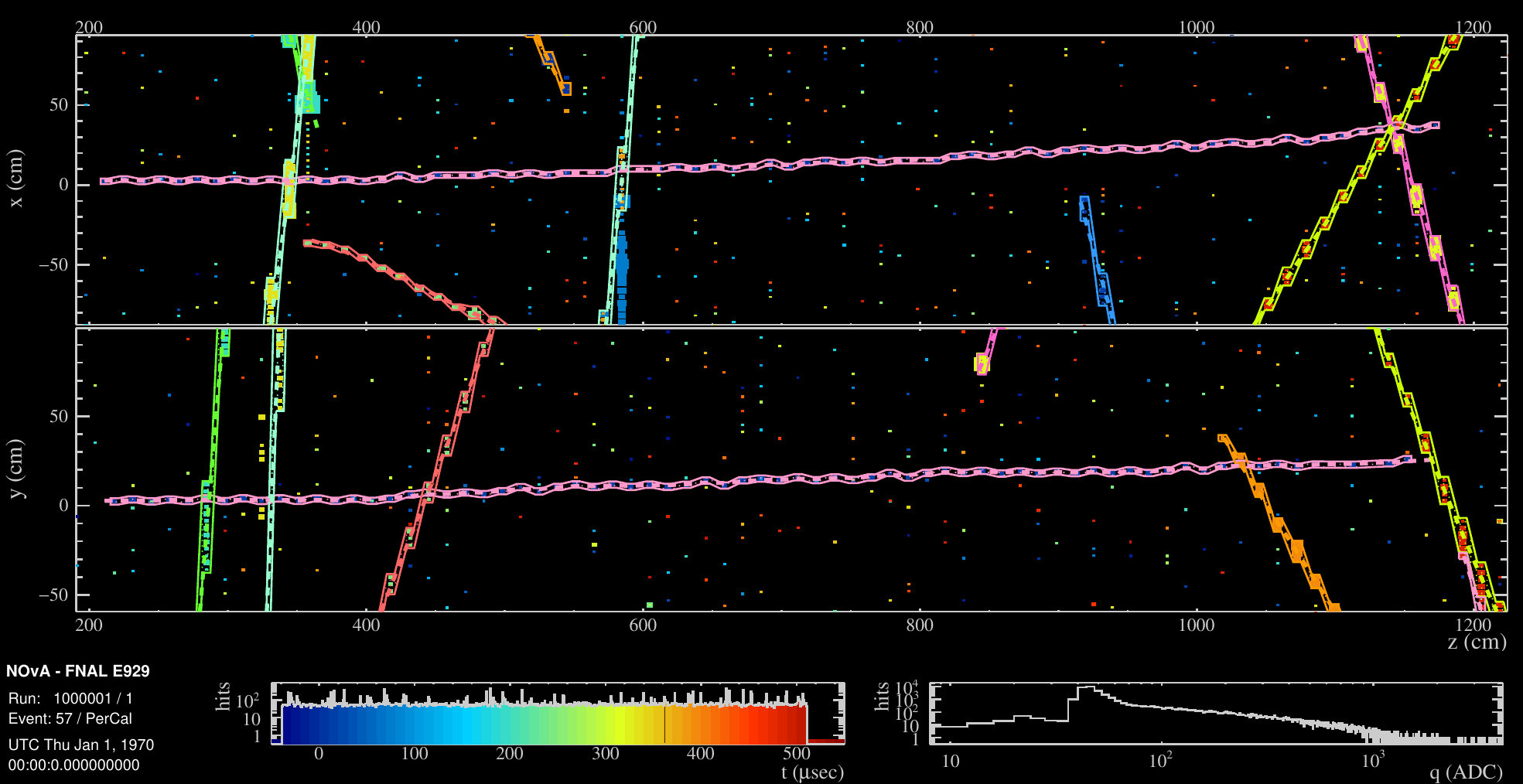}
\end{center}
\caption{\label{fig:mcmuon57} Example of a track lengthened by
cosmic overlay.}
\end{figure}

An example of a track which got 35\,cm shorter is shown in \fig{mcmuon331}.
Without overlay, two 2D tracks are formed in the $y$~view, but with the
overlay, the second one is not made and then the subsequently created 3D track
is shorter because the $x$~view track is shorter.  \fig{mcmuon57} shows a track
that got 26\,cm longer.  In this case, the last two hits in the $x$~view, which
are part of the true track, did not get included in the slice in the MC-only
file, apparently because the second to last is about 0.5\,$\mu$s out of time
with the rest.  With cosmics overlaid these two
hits did get sliced in, and also included in the track.  Two cosmics pass near
the end of the MC muon, but this appears to be incidental; they are well out of
time.  Probably a noise hit linked the trailing track hits together with the
rest of the track.

The average track became $(0.01 \pm 0.08)$\,cm shorter.
Alternatively, ignoring tracks that changed by less than 0.5\,cm, the average change was to become $(0.02 \pm 0.12)$\,cm 
shorter.  The error is statistical, and sufficiently small to show
that the impact of this effect is tiny.  Even for a rather short track
of 100\,MeV, it represents an uncertainty of 0.04\% in the extreme
case where noise or cosmics are not modeled \emph{at all}. The real 
uncertainty is at least an order of magnitude smaller.  As in the discussion 
in \sect{neutronpileup} about neutron pile-up, the effect of adding additional
hits is bigger in the Muon Catcher, but even there it is negligible.

\subsection{Muon decay}\label{sec:muondecay}

If the muon decays soon after stopping, the 1--4 hits from the Michel electron have
some chance of being reconstructed as part of the track, resulting in an overly
high track energy estimate.  This is fine as long as the data and MC agree.  
There are both physical and detector effects that could cause differences.
First, for $\mu^-$,
there is a 1\% uncertainty on the fraction of the time that the muon decays rather than undergoing nuclear capture. Second, NOvA electronics are inefficient
for recording hits in the few microseconds after a large energy
deposition in a module, and this is known to be modeled poorly.

As a back-of-the-envelope estimate, suppose that a Michel electron has a 20\%
chance of being swept up into a track if it is in time with the rest of the
track, which is reasonable given the isotropic decay direction. 
Suppose further that ``in time'' means that the muon decays in the
first 0.25\,$\mu$s, which happens 10\% of the time.  (The details of NOvA's reconstruction
algorithm means that no single number defines ``in time'', but this approximation
suffices.) A Michel goes one or two
planes and may add 15\,MeV to the track energy.  The impact is biggest on low
energy muons, so let us consider a 150\,MeV muon, for which this is a 10\%
shift.  Since it happens 2\% of the time, the mean difference if muon decay
were totally unmodeled would be 0.2\%.  

Consider $\mu^-$ capture.  There's about a 1\% absolute uncertainty on the
fraction that capture.  The systematic error on the reconstructed
muon energy from this effect is then about 0.002\%.
For $\mu^-$ that do decay, their lifetime and Michel spectrum
are a function of which isotope they are in orbit around, but this 
must have a similarly tiny impact.

Consider the electronics effects.  For them to matter, the Michel hits must be
in a module that was hit previously.  But this is unlikely if it is swept up in
the track, since to be in the same module suggests a large angle between the
muon and the Michel and to be included in the track suggests a small angle.
Suppose this  happens 10\% of the time, and that the efficiency in this case is
very badly mismodeled, by 50\%.  The contribution to the systematic error is
still something less than 0.01\%.

\subsection{Multiple scattering}\label{sec:multiple}

The reconstructed muon range is shorter than the real distance the muon travels
because the scale of multiple scattering is shorter than the NOvA cell size.  This
only matters up to discrepancies between the data and MC.

Multiple Coulomb scattering is summarized by the PDG~\cite{pdg2016} Sec.~33.3: ``it is
sufficient for many applications to use a Gaussian approximation for
the central 98\% of the projected angular distribution, with an rms width given by:''

\[\theta_0 = \frac{13.6\,\mathrm{MeV}}{\beta p} z \sqrt{x/X_0}\big [ 1 + 0.038 \log(x/X_0) \big ].\]

Nominally, \nova soup (that is, a uniform material made from stirring
the scintillator, PVC, glue, fiber and air together in the detectors' proportions)
has a radiation length, $X_0$, of 34.89\,g/cm$^2$.  Much as in \sect{elements},
modifications of the elemental composition can be tested to see how much $X_0$ can reasonably
be changed.  The effect of varying the hydrogen fraction of the oil within
reasonable bounds changes $X_0$ by only 0.01\%.  Varying the amount of
dissolved nitrogen and oxygen in the scintillator between zero and double the best
estimate makes no difference.  Changing the titanium
dioxide content of the plastic between 18.5 and 19.5 parts changes it by
0.04\%.  Finally, varying the PVC content of the plastic between 95 and 105 parts
changes it by 0.03\%.  Reasonable changes to the glue change the soup's $X_0$
by only 0.01\%.

In sum, the uncertainty on $X_0$ from elemental composition is, conservatively, about 0.05\%.
At first order, $\theta_0$ varies like the square root of the radiation length.  The bracketed
term increases this slightly.  For scattering through relevant length scales,
the uncertainty in $\theta_0$ is 0.03\%.

The ratio between the straight-line distance between a particle track's start
and end and the physical distance traversed is at most
$\sqrt{(1+\cos\theta)/2}$ for a single scatter through an angle $\theta$
halfway through the track.  At small $\theta$ this is approximately
$1-\theta^2/8$.  For a small fractional change $\Delta\theta/\theta$, the
change in the ratio is therefore $\frac14 \Delta\theta/\theta$. Given
$\Delta\theta/\theta = 0.03\%$, the error in the track length from multiple
scattering is something less than 0.007\%.

\subsection{Alignment}

The 32-plane blocks in the Far Detector are aligned relative to each other in
$x$ with an RMS somewhat less than 1\,cm.  The block-to-block
relative alignment in $y$ is expected to better than this for structural reasons.  In the
MC, random alignments in each direction are used, with a uniform distribution
of width 0.8\,cm in $x$ and 0.16\,cm in $y$.

Plane-to-plane alignment within blocks is not well understood,
although the method of assembly --- modules pushed up against a stop on top of
the block lifter --- suggests it is at the millimeter level for the edges with
the stop, in each of $x$ and $y$, and probably degrades away from that
edge.

If a muon is traveling in a straight line and crosses a block boundary for
which the alignment is mismodeled,
it will appear to bend, increasing its track
length.  Let us estimate an average mismodeling of 0.6\,cm in $x$,
which is a rough sum in quadrature of the actual alignments 
and those used in the MC.  The segment of the track at the
boundary gets $\sqrt{z^2 + x^2}/z$ longer, where $z$ is the distance between
planes in one view, 13.3\,cm, and $x$ is the MC misalignment.  This comes to
0.09\%.  In $y$, assuming 0.3\,cm typical misalignment, this is instead 0.03\%.
Assuming planes within blocks have typical alignment differences of 0.2\,cm,
each intrablock plane boundary causes the muon track to be 0.01\% too long.  The
average over a block, and therefore the detector as a whole, is a 0.014\%
overestimate of track length.

The Near Detector MC model, including the Muon Catcher, incorporates survey
information, and so has a much smaller version of this, already negligibly
small, error.  For the ratio, the errors at the Near and Far are uncorrelated,
so it is the same as the Far Detector error by itself, which is also
negligible.

\begin{table*}

{\setlength{\tabcolsep}{1.9pt}\summarytable}

\caption{\label{tab:summary} Summary of significant muon \dedx systematics.
``$\mu$C'' is the Muon Catcher.
Mass accounting is discussed in \sect{mass}. The next three columns are errors from
Bethe evaluation (\sects{sec:excitation}--\ref{sec:density}). ``Pile-up'' is
the error from neutron hits in the ND (\sect{neutronpileup}).  ``Hadronic'' is
the effective error from tracking through hadronic showers (\sect{hadronicoverlap}). The 
numbers used by the $\nu_\mu$ disappearance analysis are shown in \textbf{bold}.}

\end{table*}

\section{Conclusions}\label{sec:conclusions}

\tab{summary} summarizes the errors above and gives the totals for Near, Far,
the Muon Catcher, and the Near/Far ratios.  The 
main results are the fully correlated component of the Near and
Far uncertainties, \textbf{\sharedndfdtotalerrorsigfig}, and the uncorrelated component
of the Near/Far ratio, \textbf{\nearfartotalerrorsigfig}.

\pdffigure

In the table, errors for pairs including the Muon Catcher are given in two
parts, one for the Muon Catcher and one for the other detector (i.e. the main
ND or the FD).  This is because for dissimilar detectors it does not make sense
to assign a single number for a correlated error; an error can be fully
correlated but have an effect of a different magnitude for the two detector
components.  In each pair, the first error is for the Muon Catcher and the
second for the main ND or the FD. 

The total errors which are fully uncorrelated between detectors are:
\begin{empheq}[box=\fbox]{align}
\mathrm{ND_{uncorr}}&: \pm\nearuncorrtofartotalerror \nonumber \\
\mathrm{FD_{uncorr}}&: \pm\faruncorrtoneartotalerror \nonumber \\
\mathrm{\mu C_{uncorr}}&: \pm\mufdtotalerrormu\,\mathrm{MeV} \nonumber
\end{empheq}
The table shows, for each pair of detectors, the total error uncorrelated \emph{within} that pair, 
and the error on the ratio for that pair. 

\subsection{Gaussianity}\label{sec:gauss}

The errors are built out of several Gaussian and non-Gaussian parts by adding
each part's RMS in quadrature.  In NOvA's analysis, errors are assumed to be Gaussian, so
we have checked the Gaussianity of this error
(see \fig{nearfar}).  In cases where the underlying error
doesn't have a well-defined PDF, a rather non-Gaussian one was chosen.  Notably,
the density effect error is modeled as a uniform distribution.

The
asymmetries come primarily from the FD scintillator and ND PVC masses. In the
former case, a possibility is included that the
metering is a better measurement than the modules' volumetric measurements.  In
the latter case, it is from the offset in PVC mass due to non-uniform selection
of extrusions.  There is also an asymmetry from the boil-off
correction (see~\sect{boiloff}).

Out to $2\sigma$, a Gaussian is a reasonable approximation of each of the total
errors.  However, there is a significant long tail, such that 99.73\% of the
probability is not covered until $\sim$3.5``$\sigma$''.  This comes primarily
from the PDF chosen to represent the ICRU's evaluation
of experimental results, mostly from the 1950's and 1960's, for the mean
excitation energy $I$.  A 95\% probability is assigned for 
each to have a good error with an RMS of 0.9 of the nominal 1$\sigma$ error, and a 5\% probability of
being seriously flawed, represented as a second Gaussian with 4.5 times the
nominal error.  This (non-uniquely) satisfies the ICRU's statement of what
their errors mean.

The error on the ND/FD ratio is designed to cover several offsets --- mainly neutron pile-up ---
and as such is quite conservative when pushed past $1\sigma$.

\subsection{Relevancy to other particles}

\relevancytable

Many of the uncertainties explored in this note in the context of muons are also relevant
to other particles, as summarized in \tab{relevancy}.  The uncertainties fall into two classes,
those that affect the true \dedx and those that cause uncertainty in the muon range in some
other way.  All the major effects are in the first category.

Hadron and electron energy are measured calorimetrically, with the calorimetric
response calibrated using the minimum ionizing portion of stopping
muons.  Each of the class of errors which affects true
\dedx incurs a fully correlated error on hadrons and electron energy with muon
energy through this calibration procedure. 

The \dedx of the other particles themselves does not matter at first order,
because we don't care how many cells their energy is distributed over.  However, there
are second order effects from Birk's suppression and hit thresholds which
additionally affect non-muons.  In evaluating this, it is notable that the density effect parameterization error
affects different particles in very different amounts. Nearly all protons are
produced with energies well below where the difference in parameterizations is
largest, about 7\,GeV for protons.  For electromagnetic showers, many ionizing
electrons and positrons have energy near the maximum of the density effect
uncertainty, but a detailed study would be needed to
determine how important this energy range is overall.

All of the errors in the class that does not affect true \dedx are small. 
The details of hadronic modeling obviously affect the hadronic energy reconstruction
in the opposite way as it affects muon energy reconstruction.  Recall that the error
assigned to the muon range is for the difference in overall $\nu_\mu$ CC event
energy between data and MC.  Underneath this is the fact that the reconstructed muon, on average,
steals about one hit from the hadronic shower, so that in both data and MC, 
too much energy is assigned to the muon and too little to the hadronic system at the $\sim$2\% level.

The Coulomb correction is relevant to any charged particle produced in
a neutrino interaction.  Of course, for hadrons, the details of (strong) final state interactions
are much more uncertain, so the uncertainties surrounding the Coulomb correction
may be negligible in comparison.

While a very low energy hit from neutron capture can extend a muon track by a
whole plane or more, it only adds its very small energy to a particle with
calorimetrically reconstructed energy.  This makes pile-up less important for
non-muons.  This same reasoning applies to noise modeling,
which is already a negligible effect for muons, and even less important for other particles.

Among the other errors that are negligible for muons, muon decay uncertainties affect
\pip and K$^+$, since they decay to muons, but not to \pim or K$^-$, since they are captured by
nuclei, nor to protons.  Finally, uncertainties on multiple scattering and detector alignment
are irrelevant for particles with calorimetrically measured energies.

%
%
%
%

\nolinenumbers

\bibliographystyle{unsrt}

\begin{small} \bibliography{fnal-nova-mu-range} \end{small}

\hrule

\end{document}